\documentclass[letterpaper,english,aps,prd,preprintnumbers,nofootinbib,floatfix,onecolumn,11pt]{revtex4-2}

\usepackage{amsmath,amssymb,lmodern}
\usepackage{dcolumn}
\usepackage{graphicx,epsfig}
\usepackage{float}
\usepackage{subfig}
\usepackage{subfloat}
\usepackage[utf8]{inputenc}
\usepackage{hyperref}
\usepackage{babel}
\usepackage{bbm}

\usepackage{lipsum}
\usepackage{setspace}
\usepackage{etoolbox}
\makeatletter
\patchcmd{\@footnotetext}
  {\setspace@singlespace}{0.9}
  {}{}
\makeatother

 \newcommand{\be}{\begin{equation}}
 \newcommand{\ee}{\end{equation}}
 \newcommand{\bea}{\begin{eqnarray}}
 \newcommand{\eea}{\end{eqnarray}}

\DeclareUnicodeCharacter{00A0}{ }

\DeclareMathOperator{\io}{\iota}

\newcommand{\bo}{\bar{o}}
\newcommand{\bi}{\bar{\iota}}

\DeclareMathOperator{\brm}{\bar{\textit{m}}}
\numberwithin{equation}{section}

\newcommand{\beq}{\begin{equation}}
\newcommand{\eeq}{\end{equation}}

\def\s{\sigma}

\renewcommand*{\thefootnote}{\fnsymbol{footnote}}

\begin{document}
\count\footins = 1000

\title{Einstein-Maxwell theory and the Weyl double copy}

\author{Damien A. Easson}
\email{deasson@asu.edu}
\affiliation{Department of Physics,  
Arizona State University, Tempe, AZ 85287-1504, USA}

\author{Tucker Manton}
\email{tucker_manton@brown.edu}
\affiliation{Department of Physics, Brown University, Providence, RI 02912, USA}

\author{Andrew Svesko}
\email{a.svesko@ucl.ac.uk}
\affiliation{Department of Physics and Astronomy, University College London, London WC1E 6BT, UK}


\begin{abstract}\vspace{-2mm}
\noindent The Weyl double copy  relates vacuum solutions in general relativity to Abelian gauge fields in Minkowski spacetime. In a previous work, we showed how the Weyl double copy  can be extended to provide a treatment of external gravitational sources consistent with the classical Kerr-Schild double copy. Using this generalization, here we provide a complete double copy  analysis of electrovacuum Petrov type D spacetimes. This includes the first analysis of the charged C-metric, whose single copy interpretation invokes the two-potential formalism of electrodynamics. We also present the first double copy  prescription for the Ricci spinor, which for non-accelerating spacetimes, takes a form similar to the original double copy relation for the Weyl spinor.
\end{abstract}

\renewcommand*{\thefootnote}{\arabic{footnote}}
\setcounter{footnote}{0}

\maketitle

\newpage

\clearpage
\setcounter{page}{1}

\tableofcontents

\section{Introduction}

  Gravity  and non-Abelian gauge theories share several formal similarities. For example, general relativity and Yang-Mills are each characterized by non-linear equations of motion whose solution yields the (gauge) curvature of the background.
  Color-kinematics duality \cite{Bern:2008qj} realizes an equivalence between gravity and gauge theory at the level of perturbative scattering amplitudes. Namely, amplitudes in gravity are expressed as a double copy  of Yang-Mills gauge theory amplitudes, \emph{i.e.}, a multi-loop, multi-point gravity amplitude is given by the product of two gauge theory amplitudes \cite{Bern:2010yg,Bern:2010ue}.
  This result was anticipated by the Kawai-Lewellen-Tye (KLT) relation demonstrating any closed string amplitude may be cast as a linear sum of factors, each of which is a product of two open string amplitudes, at tree level \cite{Kawai:1985xq}. The discovery of the double copy  resulted in powerful new techniques for computing amplitudes in (super)gravity at tree-level and beyond \cite{Bern:2010ue,Bern:2010tq,Carrasco:2011mn,Oxburgh:2012zr,Bern:2013yya}, offering new insights into the fundamental nature of perturbative quantum gravity, and has numerous applications, particularly in gravitational wave physics \cite{Bern:2019nnu,Bern:2019crd,Godazgar:2020zbv,CarrilloGonzalez:2022mxx} (see \cite{Bern:2019prr,Kosower:2022yvp,Adamo:2022dcm} for comprehensive reviews).


 The success of the perturbative double copy  motivated the development of a \emph{classical} double copy, where  exact solutions in gravity have a correspondence to exact (linearized) Yang-Mills solutions \cite{Neill:2013wsa,Monteiro:2014cda,Luna:2016hge,Luna:2018dpt} (see also, \cite{Elor:2020nqe,White:2020sfn,Chacon:2021lox,Campiglia:2021srh,Adamo:2021dfg}).
 For example, the Kerr-Schild double copy  is a copying relation at the level of the graviton and gauge field \cite{Monteiro:2014cda}. Specifically, consider a spacetime $g_{\mu\nu}$ in Kerr-Schild form, 
\begin{equation}\label{KSmet}
    g_{\mu\nu}=\eta_{\mu\nu}+\phi k_\mu k_\nu, \quad g^{\mu\nu}=\eta^{\mu\nu}-\phi k^{\mu}k^{\nu},\quad g^{\mu\nu}k_\mu k_\nu = \eta^{\mu\nu}k_\mu k_\nu=0,
\end{equation}
where $\phi$ is a scalar field function of the background coordinates and $k_{\mu}$ is a null geodesic with respect to both $\eta_{\mu\nu}$ and $g_{\mu\nu}$. The Kerr-Schild ansatz (\ref{KSmet}) linearizes Einstein's field equations, and when $g_{\mu\nu}$ solves Einstein's equations, the four-vector $A_{\mu}\equiv\phi k_{\mu}$
is a solution to Maxwell's equations. Indeed, choosing $k_{0}=1$ and introducing the tensor $F_{\mu\nu}=\partial_{[\mu}A_{\nu]}$, from the trace reversed Einstein equations we identify the sourced Maxwell equations (working in units of $8\pi G=1$) \cite{Ridgway:2015fdl}
\begin{equation}\label{sourceMaxInKS}
    \begin{split}
     R^{0}_{\;\mu}=T^{0}_{\;\mu}-\frac{T}{(d-2)}\delta^{0}_{\;\mu}\longleftrightarrow\frac{1}{2}\partial^{\nu}F_{\nu\mu}=-\frac{1}{2}J_{\mu}\;,
    \end{split}
\end{equation}
where the (external) sources of the Einstein field equations correspond to (external) sources of the Maxwell equations, $J^{\mu}=-2(T^{\mu}_{\;0}-T\delta^{\mu}_{\;0}/(d-2))$, where we used the relation $R^{\mu}_{\;0}=-\frac{1}{2}\partial_{\nu}F^{\mu\nu}$.\footnote{It is generally less clear how one should interpret the spatial components  $R^{i}_{\;j}$ in the double copy . For vacuum spacetimes $R^{i}_{\;j}=0$ acts as a constraint on the Maxwell theory akin to color-kinematics duality \cite{Luna:2018dpt}, while $R^{i}_{\;j}\neq0$ is fixed by the charge density and the spatial component of the current parallel to $k_{i}$ \cite{Ridgway:2015fdl}.}

The simplest example is the four-dimensional (exterior) Schwarzschild black hole, where, under an elementary gauge transformation, $A_{\mu}$ is recognized as the Coulomb solution for a static point charge located at the origin \cite{Monteiro:2014cda};\footnote{'Vacuum' Schwarzschild is already an example of Einstein's equations with sources: $T^{\mu\nu}=Mv^{\mu}v^{\nu}\delta^{(3)}(\vec{x})$ with $v^{\mu}=(1,0,0,0)$. Correspondingly, identifying $k^{0}=+1$, the Maxwell current is $J^{\mu}=-\frac{M}{2}v^{\nu}\delta^{(3)}(\vec{x})$ \cite{Monteiro:2014cda}.} the Schwarzschild solution is the classical double copy  of a static point charge.\footnote{There is actually a bit more to this story. The Coulomb charge has been shown to double copy  into a more general static, spherically symmetric configuration known as the JNW solution \cite{Janis:1968zz}, which includes both the metric and a dilaton field \cite{Luna:2016hge,Goldberger:2016iau,Kim:2019jwm,Luna:2020adi}.} This version of the classical double copy  holds for a variety of spacetimes in higher and lower dimensions and has been generalized such that $\eta_{\mu\nu}$ in the Kerr-Schild ansatz (\ref{KSmet}) may be replaced by a curved background $\bar{g}_{\mu\nu}$, \emph{e.g.}, \cite{Luna:2015paa,Bahjat-Abbas:2017htu,Carrillo-Gonzalez:2017iyj, CarrilloGonzalez:2019gof,Keeler:2020rcv,Easson:2020esh}. Notably, the Kerr-Schild classical double copy  accounts for gravity theories with sources, such as Einstein-Maxwell theory, and is thus capable of describing, for example, charged black holes. Stronger links  between the perturbative double copy  and Kerr-Schild double copy  are presented in \cite{Guevara:2021yud,Monteiro:2021ztt}.

Another version of the classical double copy  was explored in \cite{Luna:2018dpt}, known as the Weyl double copy, and is defined as follows. Let  $\Psi_{ABCD}$ denote the completely symmetric Weyl spinor satisfying the four-dimensional vacuum Einstein equations for Petrov type D or type N spacetimes. Then the Weyl double copy relates a single copy Maxwell field strength spinor $f_{AB}$ and a (complex) scalar field $S$ to the gravity solution that constructs $\Psi_{ABCD}$,
\begin{equation}\label{OldWeylDC}
    \Psi_{ABCD}=\frac{1}{S}f_{(AB}f_{CD)}.
\end{equation}
In this context, $S$ is referred to as the zeroth copy since in the case of single Kerr-Schild spacetimes the real part of $S$ is equal to $\phi$ up to a constant. The spinors $f_{AB}$ and $\Psi_{ABCD}$ are related to their tensorial counterparts, $F_{\mu\nu}$ and $W_{\mu\nu\lambda\gamma}$, 
via Infeld-van der Waerden symbols built from Pauli matrices and spacetime vierbeins. When $\Psi_{ABCD}$ is constructed from a vacuum type D spacetime, then \cite{Luna:2018dpt}
\begin{equation}
    \Box^{(0)}S=0, \ \ \ \ \ \ \ \nabla^{(0)}_{\mu}F^{\nu\mu}=\nabla^{(0)}_\mu\tilde{F}^{\nu\mu}=0, \label{eq:WeylDCeom}\end{equation}
where $\tilde{F}_{\mu\nu}=\frac{\sqrt{-g}}{2}\varepsilon_{\mu\nu\alpha\beta}F^{\alpha\beta}$ is the dual single copy field strength tensor. The $(0)$-superscript indicates that these derivatives are taken over the flat background metric, which is obtained by taking the appropriate limit of the full spacetime metric associated to $\Psi_{ABCD}$. The Weyl double copy is consistent with the Kerr-Schild double copy and furthermore, resolves some ambiguities of the latter \cite{Luna:2018dpt}. The Weyl double copy has since been shown to hold for vacuum spacetimes of arbitrary Petrov type using methods from twistor theory \cite{White:2020sfn,Chacon:2021wbr,Chacon:2021lox}.

Despite its fundamental underpinnings, a weakness of the Weyl double copy (\ref{OldWeylDC}), compared to the Kerr-Schild double copy, is that it cannot describe classical spacetimes with external sources. More bluntly, the Weyl double copy cannot be used to construct a four-dimensional charged black hole. This is certainly a drawback to the Weyl double copy as realistic spacetimes typically involve external matter sources. Thus, to have any empirical merit, the Weyl double copy requires an extension to describe gravity theories with external sources.

 In a recent letter \cite{Easson:2021asd}, we proposed a simple generalization of the Weyl double copy (\ref{OldWeylDC}) to include sources on the gravity side of the duality. Rather than $\Psi_{ABCD}$ being constructed from a single scalar-gauge theory, we promote it to a \emph{sum} over $m$ scalar-gauge theories,
\begin{equation}\label{SourcedWeylDC}
    \Psi_{ABCD}=\sum_{n=1}^m\frac{1}{S_{(n)}}f^{(n)}_{(AB}f^{(n)}_{CD)},
\end{equation}
where for $n>1,$ each $S_{(n)}$ and $f^{(n)}_{AB}$ satisfy a particular sourced wave equation and sourced Maxwell equations respectively, 
\begin{equation}
    \nabla_\mu^{(0)}F_{(n)}^{\nu\mu}=J_{(n)}^\nu, \ \ \ \ \Box^{(0)}S_{(n)}=\rho_{(n)}^S.
\end{equation}
We thus refer to (\ref{SourcedWeylDC}) as the \textit{sourced Weyl double copy}. Here $\rho_{(n)}^{S}$ and $J^{\nu}_{(n)}$ are external sources for the zeroth\footnote{Note that here $S$ will always turn out to be time-independent. This is related to the fact we are considering stationary KS spacetimes, for which $\phi$ is time-independent. Thus, $\rho^{S}_{(n)}$ is related to the charge density $J^{0}$ of the $n^{\text{th}}$ term of the expansion \label{SourcedWeylDC}.} and single copies, respectively, with the $n=1$ term corresponding to (\ref{OldWeylDC}), or what we may consider the vacuum part of the metric. Terms associated to $n=2,3,...,m$ correspond to each external source. That is, all parameters that contribute non-trivially to the Ricci curvature pair with a gauge theory labeled by $n$ for each $n>1.$

The purpose of this article is to explore the sourced Weyl double copy for exact solutions to Einstein-Maxwell theory, focusing on stationary spacetimes. Our work not only provides the first complete double copy analysis of canonical electrovacuum type D spacetimes, including the charged C-metric and Kerr-Newman-Taub-NUT black hole, but also presents the first double copy prescription for the Ricci spinor.

The article is outlined as follows. We begin in section \ref{sec:KSandWDCLang} by offering a Lagrangian perspective of the double, single, and zeroth copy fields, together with their respective sources. Although not often discussed in the Weyl double copy literature, the Lagrangian perspective provides a useful way to organize each field and the spacetimes on which they live. This will be crucial for our purposes, since we must distinguish a Maxwell source on the double copy side from the Maxwell fields on the single copy side. We proceed by describing the relationship between the Maxwell source on the double copy side to the appropriate single copy quantity in section \ref{sec:U(1)prescription}, before presenting new relations for the spinorial counterpart of the Ricci tensor in section \ref{sec:Riccispinorscript}. 

In the remainder of the article, we apply our general prescriptions to specific electrovacuum type D spacetimes. Section \ref{sec:KerrNewman} is devoted to studying the Kerr-Newman black hole, which was given a preliminary treatment in \cite{Easson:2021asd}, however, here we include an analysis of its associated Ricci spinor. In section \ref{sec:ChargedC}, we provide the first complete analysis of the charged C-metric. Here we encounter a peculiar feature of the sourced Weyl double copy for spacetimes that are not (Riemann) asymptotically flat.\footnote{By Riemann asymptotically flat we mean that in Boyer-Lindquist coordinates, the Riemann curvature tensor vanishes in the limit of large radial coordinate. We will discuss this further in section \ref{sec:Riccispinorscript}.} The single copy theory associated to the black hole charge parameter exhibits a nontrivial divergence of the dual field strength tensor, corresponding to a magnetic current vector. A covariant description of such a solution can be obtained by introducing a so called two-potential electrodynamic theory \cite{Cabibbo:1962td},
from which we provide explicit formulae for the two necessary vector potentials associated to the charged C-metric. In section \ref{sec:gentypeD}, after we review the most general Petrov type D electrovac solution obtained by Pleba\'nski and Demia\'nski \cite{Plebanski:1976gy}, we apply the sourced Weyl double copy to the Kerr-Newman-Taub-NUT spacetime for the first time.  

Finally, we conclude in section \ref{sec:disc}, outlining future research directions. To keep this article streamlined yet self-contained, we relegate a review of the spinor formalism to Appendix \ref{app:Aspinorformalism}.

\section{Lagrangian perspective and new spinor relations}\label{sec:KSandWDCLang}


The Kerr-Schild double copy is often explained at the level of the equations of motion, \emph{e.g.}, (\ref{sourceMaxInKS}). It is illuminating to see how this translates to the actions characterizing 
the double, single, and zeroth copy fields. This is particularly useful for our purposes as we must keep track of three different Maxwell fields. Beginning with the double copy action (standard Einstein-Maxwell theory), we have a $U(1)$ gauge field minimally coupled to general relativity, 
\begin{equation}\label{dcaction}
    I_{dc}=\int d^4x\sqrt{-g}\Big(\frac{1}{2}R-\frac{1}{4}F_s^2\Big),
\end{equation}
where $F_s=dA_s$ is the field strength tensor of the Maxwell source. The single and zeroth copy theories are described by 
\begin{equation}\label{scaction}
    I_{sc}=\int d^4x\sqrt{-g^{(0)}}\Big(-\frac{1}{4}F_{(1)}^2-\frac{1}{4}F_{(2)}^2+A^{(2)}_\mu J^\mu\Big),
\end{equation}
\begin{equation}\label{zcaction}
    I_{zc}=\int d^4x\sqrt{-g^{(0)}}\Big(-\frac{1}{2}\big(\nabla^{(0)} S_{(1)}\big)^2-\frac{1}{2}\big(\nabla^{(0)} S_{(2)}\big)^2+S_{(2)} \rho^S\Big).
\end{equation}
Here $F_{(1)}$ and $S_{(1)}$ are the single copy field strength and zeroth copy scalar that map to the `vacuum' term of the Weyl tensor, $F_{(2)}$ and $S_{(2)}$ are the single and zeroth copy fields that map to the $U(1)$ charge on the gravity side, produced by $F_s$. We have explicitly introduced an electric current density $J^\mu$ directly coupling to the single copy gauge field $A_{(2)}^{\mu}$ mapped to the $U(1)$ gravity source, and, similarly, $\rho^{S}$ is the gravity source for the zeroth copy $S_{(2)}$. Further, $g_{\mu\nu}$ denotes the full spacetime metric
while $g^{(0)}_{\mu\nu}$ is the appropriate `flat space' limit of $g_{\mu\nu}$, where all parameters responsible for generating curvature are tuned to zero (we will see explicit examples of this in subsequent sections).

Varying the actions (\ref{dcaction}), (\ref{scaction}), (\ref{zcaction}) with respect to the fields $\{g^{\mu\nu},A^\mu_s,A_{(1)}^\mu,A_{(2)}^\mu,S_{(1)},S_{(2)}\}$ yields the field equations
\begin{equation}\label{EinsteinMaxEOM}
\begin{split}
    G_{\mu\nu}&=F_{s,\mu}^{\;\;\;\;\alpha}F_{s,\nu\alpha}-\frac{1}{4}g_{\mu\nu}F_s^2, \ \ \ \ \ \nabla_\nu F^{\mu\nu}_{s}=0,
\end{split}
\end{equation}
\begin{equation}\label{scEOM}
    \nabla_\nu^{(0)}F^{\mu\nu}_{(1)}=0, \ \ \ \ \ \ \nabla_\nu^{(0)}F^{\mu\nu}_{(2)}=J^\mu,
\end{equation}
\begin{equation}\label{zcEOM}
    \Box^{(0)}S_{(1)}=0, \ \ \ \ \ \ \Box^{(0)}S_{(2)}=\rho^S,
\end{equation}
where $\nabla_\mu$ is the covariant derivative with respect to $g_{\mu\nu}$ and $\nabla^{(0)}_\mu$ is the covariant derivative with respect to $g^{(0)}_{\mu\nu},$ with $\Box^{(0)}\equiv g^{\mu\nu}_{(0)}\nabla_\mu^{(0)} \nabla_\nu^{(0)}$. In principle, the scalar fields $S_{(1)}$ and $S_{(2)}$ could both be complex, however, in what follows we will see for specific examples that $S_{(1)}$ is typically complex, while $S_{(2)}$ is typically real.\footnote{For complex scalar fields, the action $I_{sc}$ should be modified by, \emph{e.g.}, replacing the real quadratic terms which their modulus, and so forth, to ensure a real action.}

The Kerr-Schild double copy \cite{Monteiro:2014cda} provides the prescription for the single copy charge density and the zeroth copy scalar charge as
\begin{equation}\label{KerrSchildsources}
    R^\mu_{ \ 0}=-\frac{1}{2} J^\mu, \ \ \ \ \ \ \ R^0_{ \ 0}=2\rho^S.
\end{equation}
Further, the actions (\ref{scEOM}) and (\ref{zcEOM}) with the sourced Weyl double copy (\ref{SourcedWeylDC}) imply the Weyl spinor for Einstein-Maxwell theories is cast as the following sum \cite{Easson:2021asd}
\begin{equation}\label{sourcedWDC}
    \Psi_{ABCD}=\frac{1}{S_{(1)}}f^{(1)}_{(AB}f^{(1)}_{CD)}+\frac{1}{S_{(2)}}f^{(2)}_{(AB}f^{(2)}_{CD)}.
\end{equation}
The spinors $\Psi_{ABCD}$, $f^{(1)}_{AB}$, and $f^{(2)}_{AB}$ are related to their tensorial counterparts through contractions with the Infeld-van der Waerden symbols $\s^{\mu\nu}_{AB}$, constructed from Pauli matrices and spacetime vierbeins (see Appendix \ref{app:Aspinorformalism}). The Weyl tensor $W_{\mu\nu\alpha\beta}$ is related to the (completely symmetric) Weyl spinor $\Psi_{ABCD}$ by
\begin{equation}
    \Psi_{ABCD}=\frac{1}{4}W_{\mu\nu\alpha\beta}\s^{\mu\nu}_{AB}\s^{\alpha\beta}_{CD}.
\end{equation}
Similarly, the source field strength spinor is 
\begin{equation}\label{fs}
    f^s_{AB}=\frac{1}{2}F^s_{\mu\nu}\s^{\mu\nu}_{AB},
\end{equation}
while the single copy field strengths are
\begin{equation}\label{thefns}
    f^{(n)}_{AB}=\frac{1}{2}F^{(n)}_{\mu\nu}\s^{(0),\mu\nu}_{AB}, \ \ \ \ \ \ n=1,2.
\end{equation}
We write $\s^{(0),\mu\nu}_{AB}$ to make explicit that the vierbeins required to construct (\ref{thefns}) are those which build the appropriate flat limit of the full spacetime metric. 

Due to the presence of $F^s_{\mu\nu},$ the solution necessarily has a nonzero Ricci tensor and associated Ricci spinor. The key result from \cite{Easson:2021asd} is that by writing the Weyl spinor as the sum of two terms as (\ref{sourcedWDC}), then the field equations (\ref{scEOM}) and (\ref{zcEOM}) are satisfied and the sources are given by (\ref{KerrSchildsources}). We therefore have three quantities which require a single and zeroth copy prescription, namely,
\begin{equation}
    \Psi_{ABCD}, \ \ \ \ \Phi_{ABC'D'}, \ \ \ \ \ f^s_{AB},
\end{equation}
where $\Phi_{ABC'D'}$ is related to the Ricci tensor and will be defined shortly (see Appendix \ref{app:Aspinorformalism} for a detailed discussion).  
Interestingly, for all of the electrovacuum spacetimes considered in this article
we find 
\begin{equation}\label{foverS}
    \frac{1}{S_{(1)}}f^{(1)}_{AB}\propto  \frac{1}{S_{(2)}}f^{(2)}_{AB}.
\end{equation}
Here the proportionality symbol denotes equality up to a constant ratio depending on the dynamical parameters of the full spacetime.  As noted, the sourced Weyl double copy (\ref{sourcedWDC}) provides the prescription for $\Psi_{ABCD}.$ We now provide copying prescriptions for the double copy field strength and Ricci spinors, $f^s_{AB}$ and $\Phi_{ABC'D'},$ respectively.


\subsection{A prescription for the $U(1)$ source}\label{sec:U(1)prescription}

The field equation for $F^s_{\mu\nu}$ (\ref{EinsteinMaxEOM}) is that of a free Maxwell gauge field over the full metric $g_{\mu\nu},$ while the field equation for $F^{(1)}_{\mu\nu}$ in (\ref{scEOM}) is that of a free Maxwell field over $g^{(0)}_{\mu\nu}.$ Since they both satisfy a vacuum equation, the form of their spinor field strengths are in fact equivalent, which can be seen as follows. Define the frame field strength associated to $F^{(1)}_{\mu\nu}$ as $F^{(1)}_{ab}=F^{(1)}_{\mu\nu}(e^{(0)})^\mu_a(e^{(0)})^\nu_b$. It is straightforward to show when $F^{(1)}_{\mu\nu}$ satisfies a vacuum equation over $g^{(0)}_{\mu\nu},$ then $F^{(1)}_{ab}$ satisfies the vacuum equations over $\eta_{ab}=(-1,1,1,1).$ Explicitly, in terms of vierbeins and Infeld-van der Waerden symbols (\ref{IvdWsymbols}),  
\begin{equation}\label{f1frameF}
    \begin{split}
        f^{(1)}_{AB}&=\frac{1}{2}F^{(1)}_{\mu\nu}(e^{(0)})^\mu_a(e^{(0)})^\nu_b\sigma^{[a}_{AC'}\bar{\sigma}^{b]C'}_{ \ \ B} 
        =\frac{1}{2}F^{(1)}_{ab}\sigma^{[a}_{AC'}\bar{\sigma}^{b]C'}_{ \ \ B},
    \end{split}
\end{equation}
where we used $(e^{(0)})^a_\mu(e^{(0)})^\mu_b=\delta^a_b.$ For the source field strength, we have
\begin{equation}\label{fsframeF}
    \begin{split}
        f^{s}_{AB}&=\frac{1}{2}F^{s}_{\mu\nu}e^\mu_ae^\nu_b\sigma^{[a}_{AC'}\bar{\sigma}^{b]C'}_{ \ \ B} 
        =\frac{1}{2}F^{s}_{ab}\sigma^{[a}_{AC'}\bar{\sigma}^{b]C'}_{ \ \ B},
    \end{split}
\end{equation}
where $F^s_{ab}=F^s_{\mu\nu}e^\mu_ae^\nu_b$ also satisfies the vacuum Maxwell equation over $\eta_{ab},$\footnote{To see this, note that $\nabla_a F^{ab}=e^\lambda_a(\nabla_\lambda F^{\mu\nu})e_\mu^a e_\nu^b=\delta^\lambda_\mu(\nabla_\lambda F^{\mu\nu})e_\nu^b=(\nabla_\mu F^{\mu\nu})e^b_\nu=0$, where we have used the no-torsion constraint $\nabla_\lambda e^a_\nu=0$.} and we used $e^a_\mu e^\mu_b=\delta^a_b.$ The key point is that although $F^s_{\mu\nu}$ and $F^{(1)}_{\mu\nu}$ are constructed using different vierbeins, their frame field strengths $F^s_{ab}$ and $F^{(1)}_{ab}$ both satisfy the vacuum Maxwell equations over $\eta_{ab}.$ Consequently,
\begin{equation}
    F^s_{ab}\propto F^{(1)}_{ab} \ \ \ \Leftrightarrow \ \ \ f^{s}_{AB}\propto f^{(1)}_{AB},
\label{eq:FsF1}\end{equation}
which is clear from the rightmost equality in (\ref{f1frameF}) and (\ref{fsframeF}).


\subsection{A prescription for the Ricci spinor}\label{sec:Riccispinorscript}

We describe the essential relations between the Ricci tensor $R_{\mu\nu}$ and the Ricci spinor $\Phi_{ABC'D'}=\Phi_{(AB)(C'D')}$ in Appendix \ref{app:Aspinorformalism}. For Einstein-Maxwell theory, the only nonzero contraction\footnote{More specifically, we work with the principle null tetrad that results in the Weyl scalar $\Psi_2\neq0,$ while all others (\ref{PsiIs}) vanish. The same tetrad is used to compute $\Phi_{ABC'D'}$, which is such that only $\Phi_{11}\neq0.$ Choosing a different tetrad generically will result in additional Ricci scalars being nonzero.} from (\ref{RicciComps}) is the component $\Phi_{010'1'}\equiv \Phi_{11}=-\frac{1}{2}R_{\mu\nu}n^\mu l^\nu+3\Pi,$ where $\Pi=\frac{1}{12}R$. The Ricci spinor thus dramatically simplifies to
\begin{equation}\label{PhiEM}
    \Phi^{EM}_{ABC'D'}=4\Phi_{11}o_{(A}\iota_{B)}\bo_{(C'}\bi_{D')},
\end{equation}
Now, the fields $S_{(2)}$ and $f^{(2)}_{AB}$ are nonzero due to the presence of $F^s_{\mu\nu},$ which in turn results in $\Phi_{ABC'D'}$ being nonzero. Therefore we expect $\Phi_{ABC'D'}$ to be related to the $S_{(2)}$ and $f^{(2)}_{AB}.$
The spinor $\Phi_{ABC'D'}$ clearly carries two conjugate indices, and itself is real for all spacetimes we consider. In general, $f^{(2)}_{AB}$ is complex, therefore its conjugate must be involved in the copying prescription for $\Phi_{ABC'D'}$. In fact, the desired relationship is analogous to the prescription for the Weyl spinor,
\begin{equation}\label{flatPhi}
    \Phi_{ABC'D'}=\frac{1}{3} \frac{1}{S_{(2)}}f^{(2)}_{AB}\bar{f}^{(2)}_{C'D'},
\end{equation}
where $\bar{f}^{(2)}_{A'B'}\propto \bo_{(A'}\bi_{B')}$ is the complex conjugate of $f^{(2)}_{AB}$.

As we stated previously, the scalar and spinor field strength for the two terms in the sourced Weyl double copy (\ref{sourcedWDC}) are proportional up to parameters characterizing the spacetime when we take the combination $\frac{1}{S}f_{AB}$, (\ref{foverS}). Further, when the scalar fields $S_{(1)}$ and $S_{(2)}$ obey
\begin{equation}\label{flatSs}
S_{(2)}\propto |S_{(1)}|^2, 
\end{equation}
then, combined with (\ref{foverS}) it follows
\begin{equation}
    \frac{1}{S_{(2)}}f^{(2)}_{AB}\bar{f}^{(2)}_{C'D'}\propto\Bigg(\frac{1}{S_{(1)}}f^{(1)}_{AB}\Bigg)\Bigg(\frac{S_{(2)}}{S_{(1)}}f^{(1)}_{CD}\Bigg)^{\hspace{-1mm}\ast}=\frac{S_{(2)}}{|S_{(1)}|^2}f^{(1)}_{AB}\bar{f}^{(1)}_{C'D'}\propto f^{(1)}_{AB}\bar{f}^{(1)}_{C'D'},
\label{eq:S2f2f1rel}\end{equation}
where the $\ast$ refers to complex conjugation. Subsequently,
\begin{equation}\label{nonflatPhi}
    \Phi_{ABC'D'}\propto f^{(1)}_{AB}\bar{f}^{(1)}_{C'D'}.
\end{equation}
Despite relating the Ricci spinor -- which vanishes in pure vacuum -- directly to the product of the single copy field strength spinors associated with the vacuum solution, this relationship in fact should not come as a surprise. In Einstein-Maxwell theory, the trace reversed Einstein equations yield $R_{\mu\nu}=T_{\mu\nu}$. In the language of spinors, this is \cite{Penrose:1985bww} 
\begin{equation}\label{EMaxinspinor}
    \Phi_{ABC'D'}\propto f^s_{AB}\bar{f}^s_{C'D'}.
\end{equation}
As we showed in section \ref{sec:U(1)prescription}, $f^{(1)}_{AB}\propto f^s_{AB}$, therefore (\ref{EMaxinspinor}) implies (\ref{nonflatPhi}).

\vspace{.5cm} 

\noindent \textbf{A comment on asymptotic flatness}

The single copy fields building the Ricci spinor (\ref{PhiEM}) will be different depending on the behavior of the curvature of the double copy spacetime at asymptotic spatial infinity. In particular and as we will show explicitly in subsequent sections, \emph{all} spacetimes examined in this article, including the most general type D spacetime, naturally have Ricci spinor obeying (\ref{EMaxinspinor}) and therefore (\ref{nonflatPhi}); however, not all electrovacuum spacetimes satisfy the relations (\ref{flatPhi}) or (\ref{flatSs}).

To appreciate this point, we briefly comment on the notion of asymptotic flatness. A traditional and practical definition is that a spacetime is said to be asymptotically flat if at spatial infinity the spacetime metric takes a Minkowski form. For example, the Schwarzschild black hole is asymptotically flat while the Taub-NUT solution is not, the latter having a non-zero $g_{t\phi}$ component in the large radial $r$ limit in Boyer-Lindquist coordinates. With this notion, any metric which is asymptotically flat will necessarily have vanishing curvature at spatial infinity. However, the reverse need not be true: asymptotically, all components of the Riemann curvature of the Taub-NUT solution vanish. In what follows, with a slight abuse of terminology, we will work with a refined notion of asymptotically flat, dubbed `Riemann asymptotically flat',\footnote{In other words, asymptotic flatness implies Riemann asymptotic flatness, but the converse need not be true.} as being a spacetime whose curvature tensor vanishes at infinity. By this notion, the (charged) Taub-NUT solution is Riemann asymptotically flat (counter to the traditional sense of asymptotic flatness of the metric).

Intriguingly, when the Riemann tensor is non-vanishing at infinity, we will observe the gauge field strengths $f^{(2)}_{AB}$ do not satisfy the relations (\ref{flatPhi}) or (\ref{flatSs}). In particular, this is the case for the charged C-metric and general Plebanski and Demianski solution \cite{Plebanski:1976gy}, for which the charged C-metric is a special limit. However, the Kerr-Newmann-Taub-NUT does obey these relations, and, subsequently, the Ricci spinor may be cast as (\ref{flatPhi}). 

Further, as we will see, for the two non-Riemann asymptotically flat spacetimes we analyze, the dual field strength tensor $\tilde{F}^{(2)}_{\mu\nu}$ associated with $f^{(2)}_{AB}$ has a nontrivial divergence, signaling the presence of a magnetic four-current;
\begin{equation}
    \nabla^{(0)}_\nu \tilde{F}_{(2)}^{\mu\nu}=J^\mu_{m}.
\end{equation}
This is equivalent to the field strength tensor failing to satisfy the Bianchi identity $\nabla^{(0)}_{[\mu}F_{\rho\sigma]}\neq 0.$ Traditional Maxwell theory cannot accommodate a nontrivial divergence of the dual field strength. However, a magnetic current can be accounted for if Maxwell theory is extended to include a second vector potential \cite{Dirac:1948um,Cabibbo:1962td}, which we will describe in section \ref{sec:ChargedC}.

Having established a general prescription relating the double copy spinor fields to single and zeroth copies, let us survey several electrovacuum type D spacetimes. We will find the sum over products of gauge field strengths (\ref{sourcedWDC}) leads to a consistent Weyl double copy that incorporates external Maxwell sources.


\section{Kerr-Newman black hole}\label{sec:KerrNewman}

The Kerr-Newman solution describes a rotating black hole with rotation parameter $a$ that carries an electric charge $Q$, and is an exact solution to the Einstein-Maxwell equations. The line element in Boyer-Lindquist coordinates is
\begin{equation}\label{KerrNewmanMet}
\begin{split}
    &ds^2= -\frac{\Delta }{\rho^2}(dt-a\sin^2\theta d\phi)^2 +\frac{\sin^2\theta }{\rho^2}\Big((r^2+a^2)d\phi-adt\Big)^2 +\frac{\rho^2}{\Delta} dr^2+\rho^2d\theta^2,\\
    & \rho^2=r^2+a^2\cos^2\theta, \ \ \ \ \ \ \ \Delta = r^2-2Mr+a^2+Q^2.
\end{split}
\end{equation}
The flat space limit of the line element is Minkowski space in oblate spheroidal coordinates,
\begin{equation}\label{OblateSpheroid}
    ds_{(0)}^2=-dt^2+\frac{\rho^2}{r^2+a^2}dr^2+\rho^2d\theta^2+(a^2+r^2)\sin^2\theta d\phi^2,
\end{equation}
which is obtained by tuning the `dynamical parameters' $\{M\rightarrow 0, Q\rightarrow 0 \}$, as this is the limit in which the Weyl curvature is vanishing. 

The Weyl spinor associated to the Kerr-Newman solution is given by \cite{Adamo:2014baa}
\begin{equation}
\begin{split} 
    \Psi_{ABCD}&=\Psi^{(1)}_{ABCD}+\Psi^{(2)}_{ABCD} \\
    &=6\Bigg(-\frac{M}{(r+ia\cos\theta)^3}+\frac{Q^2}{(r+ia\cos\theta)^3(r-ia\cos\theta)}\Bigg)o_{(A}o_B\iota_C\iota_{D)}.
    \end{split} 
\label{eq:WeylspinorKN}\end{equation}
The contribution proportional to parameter $M$ is present in the `pure' Weyl double copy
and appears in the first term in the sourced Weyl double copy sum (\ref{sourcedWDC})
\begin{equation}
    \frac{1}{S_{(1)}}f^{(1)}_{(AB}f^{(1)}_{CD)}=-\frac{6M}{(r+ia\cos\theta)^3}o_{(A}o_B\iota_C\iota_{D)}.
\end{equation}
The single and zeroth copy interpretation of this quantity was given in \cite{Luna:2018dpt}, where the scalar and gauge fields were identified to be
\begin{equation}\label{S1f1KerrNewman}
    S_{(1)}=-\frac{q_1^2}{6M}\frac{1}{r+ia\cos\theta}, \ \ \ \ \ \ \ \ \ \ \ \ \  f^{(1)}_{AB}=\frac{q_1}{(r+ia\cos\theta)^2}o_{(A}\iota_{B)},
\end{equation}
both of which satisfy the vacuum equations over the flat space metric (\ref{OblateSpheroid}). Here one has introduced the free parameter $q_{1}$ associated with the mass parameter $M$ \cite{Monteiro:2014cda}.\footnote{Note the real scalar $S_{(1)}+\bar{S}_{(1)}$ is identified as the Kerr-Schild function $\phi$ in the metric decomposition \ref{KSmet}.}

\vspace{.7cm}

\noindent \textbf{Sourced Weyl double copy}

The $Q^{2}$ contribution to the Weyl spinor (\ref{eq:WeylspinorKN}) implies the second term in the sum (\ref{sourcedWDC}) is 
\begin{equation}
    \frac{1}{S_{(2)}}f^{(2)}_{(AB}f^{(2)}_{CD)}=\Bigg(\frac{6Q^2}{(r+ia\cos\theta)^3(r-ia\cos\theta)}\Bigg)o_{(A}o_B\iota_C\iota_{D)}.
\end{equation}
Following \cite{Easson:2021asd}, we choose the zeroth copy scalar field and single copy spinor field strength corresponding to the Maxwell source of the full space to be
\begin{equation}\label{S2f2KerrNewman}
    S_{(2)}=\frac{q_2^2}{6Q^2}\frac{1}{(r+ia\cos\theta)(r-ia\cos\theta)}, \ \ \ \ \ \ \ \ \ f^{(2)}_{AB} = \frac{q_2}{(r+ia\cos\theta)^2(r-ia\cos\theta)}o_{(A}\iota_{B)},
\end{equation}
where we have introduced a second free parameter $q_{2}$ associated with the black hole charge $Q$. With the respect to the flat space metric (\ref{OblateSpheroid}), it is straightforward to show the zeroth copy scalar field satisfies
\begin{equation}
    \Box^{(0)}S_{(2)}=-\frac{2q_2^2}{Q^2}\frac{(r^2+a^2)+a^2\sin\theta^2}{\rho^6},
\label{eq:scalarwaveeomKN}\end{equation}
while the field strength obeys
\begin{equation}
    \nabla^{(0)}_{\mu}F_{(2)}^{\nu\mu}=-q_2\begin{pmatrix}
    \frac{(r^2+a^2)+a^2\sin\theta^2}{2\rho^6} \\
    0 \\
    0 \\
    \frac{a}{\rho^6}
    \end{pmatrix}, \ \ \ \ \ \ \ \ \ \ \ \nabla^{(0)}_\mu \tilde{F}_{(2)}^{\nu\mu}=0.
\end{equation}
The presence of the rotation parameter introduces a current density and therefore a magnetic field, as expected. Another new element not present in the pure Weyl double copy is the gravitational energy density $\rho^{\text{grav}}$,
\begin{equation}
    \rho^{\text{grav}} = 8Q^2\frac{(r^2+a^2)+a^2\sin\theta^2}{\rho^6}.
\end{equation}
Comparing to the scalar wave equations (\ref{eq:scalarwaveeomKN}), the gravitational energy density is proportional to the single copy electric charge density $\rho^{e}\equiv J^{0}$, and the scalar charge density $\rho^{S}$.  

Computing the $(1,1)$ Ricci tensor, we find
\begin{equation}
    R^\mu_{ \ 0}=-16Q^2 \begin{pmatrix} \frac{(r^2+a^2)+a^2\sin^2\theta }{2\rho^6} \\ 0 \\ 0 \\ \frac{a}{\rho^6}
    \end{pmatrix}\propto J^\mu_{(2)}, 
\label{eq:RiccimixedKN}\end{equation}
in agreement with our expectations from the sourced Kerr-Schild double copy. By inspection of (\ref{S1f1KerrNewman}) and (\ref{S2f2KerrNewman}), we see
\begin{equation}\label{12relations}
    \frac{1}{S_{(1)}}f^{(1)}_{AB}\propto\frac{1}{S_{(2)}}f^{(2)}_{AB}, \ \ \ \ \ \ \ \text{and} \ \ \ \ \ \ \ \ S_{(2)}\propto |S_{(1)}|^2,
\end{equation}
as claimed above for asymptotically flat spacetimes. We will return to this point momentarily. 

As discussed in \cite{Easson:2021asd}, the gauge field required to supplement the Kerr-Newman solution is
\begin{equation}
    A_\mu = \frac{Q^2r}{\rho^2}\Big(1,\rho^{2}/(r^{2}+a^{2}),0,a\sin^2\theta\Big),
\end{equation}
which we note is (gauge) equivalent to the single copy gauge field associated to the Kerr solution, i.e. the solution associated to $f^{(1)}_{AB}.$ More precisely, when the Kerr-Newman black hole is expressed in Kerr-Schild form (\ref{KSmet}), one may decompose the Kerr-Schild scalar $\phi$ as
\beq \phi=\phi^{K}+\phi^{N}\;,\quad \phi^{K}=\frac{2Mr}{\rho^{2}}\;,\quad \phi^{N}=-\frac{Q^{2}}{\rho^{2}}\;.\eeq

In this case, the single copy gauge field $A_{\mu}\equiv\phi k_{\mu}$, for null vector $k_{\mu}=(1,\rho^{2}/(r^{2}+a^{2}),0,a\sin^{2}\theta)$, takes a similar decomposition, $A_{\mu}=A^{K}_{\mu}+A^{N}_{\mu}$. The gauge field $A_{\mu}^{K}$ matches the one found in \cite{Luna:2018dpt} (up to a parameter rescaling), and whose corresponding gauge field strength $F^{(1)}_{\mu\nu}$ is divergenceless. Meanwhile, the gauge field $A_{\mu}^{N}$ produces a field strength $F^{(2)}_{\mu\nu}$ whose divergence is proportional to (\ref{eq:RiccimixedKN}). This demonstrates a consistency between the Kerr-Schild and sourced Weyl double copy.


\vspace{.5cm}

 \noindent \textbf{Ricci spinor}

We now analyze the Ricci spinor for the Kerr-Newman solution and provide its double copy analysis. 
The only non-zero component is $\Phi_{010'1'}$ such that \cite{Podolsky:2006px}
\begin{equation}\label{Riccispinorterm}
   \Phi_{ABC'D'}=\frac{2Q^2}{(r^2+a^2\cos^2\theta)^2}o_{(A}\iota_{B)}\bo_{(C'}\bi_{D')}.
\end{equation}
Comparing to the single copy field strength spinor $f_{AB}^{(1)}$ (\ref{S1f1KerrNewman}), we see the Ricci spinor is proportional to the complex square of $f_{AB}^{(1)}$ 
\beq f^{(1)}_{AB}\bar{f}^{(1)}_{C'D'}=\frac{q_{1}^{2}}{(r^{2}+a^{2}\cos\theta^{2})^{2}}o_{(A}\iota_{B)}\bo_{(C'}\bi_{D')}=\frac{q_{1}^{2}}{2Q^{2}}\Phi_{ABC'D'}\;.\label{eq:kerrnewriccif1}\eeq
Alternately, via the zeroth and single copy fields $S^{(2)}$ and $f^{(2)}_{AB}$ (\ref{S2f2KerrNewman}), we observe 
\begin{equation}\label{RicciDC}
    \frac{1}{S_{(2)}}f^{(2)}_{AB}\bar{f}^{(2)}_{C'D'}=\frac{6Q^2}{(r^2+a^2\cos^2\theta)^2}o_{(A}\iota_{B)}\bar{o}_{(C'}\bar{\iota}_{D')}=3 \Phi_{ABC'D'}.
\end{equation}
Hence, the Ricci spinor for the Kerr-Newman black hole may be characterized by the product of a single copy field strength spinor and its complex conjugate, reminiscent of the double copy relation of the Weyl spinor. The fact the Ricci spinor can be cast as a product of $f^{(2)}_{AB}$ is the relation which is only enjoyed by Riemann asymptotically flat spacetimes,
as we now explore.\footnote{The double copy relations of the Reissner-Nordstr\"om black hole follow from the above analysis of the Kerr-Newman black hole in the $a\rightarrow 0$ limit, or equivalently in the limit that $\alpha\rightarrow 0$ of the charged C-metric of section \ref{sec:ChargedC}. }


\section{Charged C-metric}\label{sec:ChargedC}

The charged C-metric will serve as a simple example of the rich new properties of non-Riemann asymptotically flat, non-vacuum spacetimes in the framework of the Weyl double copy. In spherical-like coordinates, its metric takes the form \cite{Podolsky:2006px}
\begin{equation}\label{fullmet}
    ds^2=\frac{1}{\Omega^2}\Big[-\frac{A}{r^2}dt^2+\frac{r^2}{A}dr^2+\frac{r^2}{B}d\theta^2+Br^2\sin^2\theta d\phi^2\Big],
\end{equation}
with 
\begin{equation}
    \begin{split}
        \Omega &=1-\alpha r\cos\theta, \\
        A&=(Q^2-2Mr+r^2)(1-\alpha^2r^2), \\
        B&=1-2\alpha M\cos\theta+Q^2\alpha^2\cos^2\theta.
    \end{split}
\end{equation}
This solution describes a (charged $Q$) black hole of mass $M$ that is uniformly accelerating with acceleration parameter $\alpha$, generated by a cosmic string pulling it outward. Alternatively, the solution represents a pair of charged, uniformly accelerating black holes moving in opposite directions, either due to a connecting cosmic string between pushing them apart or as two semi-infinite cosmic strings pulling them away from each other (cf.  \cite{Griffiths:2006tk}). In the $\alpha\to0$ limit, the solution describes a single Reissner-Nordstr\"om black hole.
Importantly, the C-metric is not asymptotically flat, in the sense of the metric not being asymptotically Minkowskian and, correspondingly, has non-vanishing Riemann curvature at infinity.

With respect to these coordinates, the Weyl spinor $\Psi_{ABCD}$ naturally decomposes into a term proportional to $M$ and $Q^{2}$:
\begin{equation}\label{CmetWeyl}
    \begin{split}
        \Psi_{ABCD}&=\Psi^{(1)}_{ABCD}+\Psi^{(2)}_{ABCD} \\
        &=-6M\frac{(1-\alpha r\cos\theta)^3}{r^3}o_{(A}o_B\iota_C\iota_{D)}+6Q^2\frac{(1+\alpha r\cos\theta)(1-\alpha  r\cos\theta)^3}{r^4}o_{(A}o_B\iota_C\iota_{D)},
    \end{split}
\end{equation}
while the Ricci spinor and Ricci tensor are, respectively,
\begin{equation}\label{RicSpin}
    \Phi_{ABC'D'}=2Q^2\frac{(1-\alpha r\cos\theta)^4}{r^4}o_{(A}\iota_{B)}\bar{o}_{(C'}\bar{\iota}_{D')},
\end{equation}
\begin{equation}\label{Cricci}
     R^\mu_{ \ \nu}=Q^2\frac{(1-\alpha r\cos\theta)^4}{r^4}\text{diag}\big(-1,-1,1,1\big).
\end{equation}



The pure Weyl double copy of the neutral C-metric, with Weyl spinor $\Psi_{ABCD}^{(1)}$, was examined in  \cite{Luna:2018dpt}, where it was recognized that the gauge potential $A_{\mu}$ associated to the single copy field strength $F^{(1)}_{\mu\nu}$ corresponds to the Li\'enard-Wiechert potential describing a pair of causally disconnected charges, uniformly accelerating in opposite directions with acceleration $\alpha$. The zeroth and single copy fields $S_{(1)}$ and $f^{(1)}_{AB}$ associated with $\Psi^{(1)}_{ABCD}$ in (\ref{CmetWeyl}) are
\begin{equation}
    S_{(1)}=\frac{q_1^2}{M}\frac{1-\alpha r\cos\theta}{r}, \ \ \ \ \ \ f^{(1)}_{AB}=q_1\frac{(1-\alpha r\cos\theta)^2}{r^2}o_{(A}\iota_{B)},
\end{equation}
where we have introduced the real, free parameter $q_{1}$, understood as the charge of the particle in the single copy picture. These fields construct the product $\Psi^{(1)}_{ABCD}=\frac{1}{S_{(1)}}f^{(1)}_{(AB}f^{(1)}_{CD)}$ and live over the flat background obtained by sending $\{M,Q\}\rightarrow\{0,0\}$ in (\ref{fullmet}),
\begin{equation}\label{AccelMink}
    ds_{(0)}^2=\frac{1}{\Omega^2}\Big[-(1-\alpha^2r^2)dt^2+\frac{dr^2}{1-\alpha^2r^2}+r^2d\Omega^2\Big],
\end{equation}
which is Minkowski space in accelerated coordinates. Additionally, the zeroth and single copy field strength tensor associated to $f^{(1)}_{AB}$ satisfy the vacuum equations
\begin{equation}
    \Box^{(0)}S_{(1)}=0, \ \ \ \ \ \nabla^{(0)}_{\nu}F^{\mu\nu}_{(1)}=\nabla^{(0)}_\nu\tilde{F}^{\mu\nu}_{(1)}=0.
\end{equation}
From here, it is straightforward to show the gauge potential corresponding to $F^{(1)}_{\mu\nu}$ is
\begin{equation}\label{A1chargedC}
    A_{(1)}=\frac{q_1}{r}dt,
\end{equation}
which (functionally) agrees with the potential required to solve the Einstein-Maxwell equations on the double copy side, $A^s=\frac{Q}{r}dt.$ To express the gauge potential in a more standard Li\'enard-Wiechert form, one may transform the accelerated coordinate system to Minkowski space in Cartesian coordinates \cite{Luna:2018dpt}.

\vspace{.5cm}

\noindent \textbf{Sourced Weyl double copy}

Let us now determine the fields required to construct the second term $\Psi^{(2)}_{ABCD}$ of the complete Weyl spinor (\ref{CmetWeyl}) such that it may be cast as
\begin{equation}
\Psi^{(2)}_{ABCD}=
\frac{1}{S_{(2)}}f^{(2)}_{(AB}f^{(2)}_{CD)}.
\end{equation}
The scalar field is required to satisfy $\Box^{(0)}S_{(2)}\propto R^t_{ \ t}\sim (1-\alpha r\cos\theta)^4/r^4,$ which has the solution
\begin{equation}
    S_{(2)}=\frac{q_2^2}{6Q^2}\frac{(1+\alpha r\cos\theta)(1-\alpha r\cos\theta)}{r^2},
\label{eq:S2Cmet}\end{equation}
where we have introduced another free parameter $q_2.$ Thus, the single copy field strength is
\begin{equation}
    f^{(2)}_{AB}=q_2\frac{(1+\alpha r\cos\theta)(1-\alpha r\cos\theta)^2}{r^3}o_{(A}\iota_{B)}.
\label{eq:f2Cmet}\end{equation} 
Correspondingly, the tensor field strength associated to $f^{(2)}_{AB}$ is
\begin{equation}
    F^{(2)}=-q_2\frac{1+\alpha r\cos\theta}{2r^3}dt\wedge dr.
\label{eq:F2Cmet}\end{equation}
whose divergence corresponds to the expectation from the Kerr-Schild double copy,
\begin{equation}\label{dF}
    \nabla^{(0)}_{\nu}F^{\mu\nu}_{(2)}=-\frac{q_2}{2}\frac{(1-\alpha r\cos\theta)^4}{r^4}\delta^\mu_t \equiv J^\mu_e\propto R^\mu_{ \ t},
\end{equation}
A new feature of the C-metric is that we also have a nontrivial divergence of the dual tensor $\tilde{F}^{\mu\nu}_{(2)}$,
\begin{equation}\label{dFtilde}
    \nabla^{(0)}_\nu \tilde{F}^{\mu\nu}_{(2)}=-\frac{q_2\alpha}{2}\frac{(1-\alpha r\cos\theta)^4}{r^4}\delta^t_\phi\equiv J_m^\mu.
\end{equation}
A non-zero divergence of the dual tensor typically implies the presence of a magnetic monopole with its own magnetic current, $J_{m}^{\mu}$. We emphasize, this magnetic current is not present in the neutral C-metric. To better interpret the corresponding single copy field strength $F^{(2)}_{\mu\nu}$, let us briefly digress to discuss how two-potential electrodynamics can account for the new magnetic current. 




\vspace{.5cm}

\noindent \textbf{Two-potential electrodynamics}

In the presence of magnetic charges, Maxwell's equations may be modified whilst maintaining the traditional definition of the Maxwell potential $A_{\mu}$. Doing so, however, comes at the cost of placing magnetic charges at the end of a string on which $A_{\mu}$ develops a singularity \cite{Dirac:1931kp,Dirac:1948um}. Alternatively, one may avoid introducing these singular and non-local Dirac strings by way of two-potential electrodynamics, where one introduces a pseudo-4-vector potential, $C_{\mu}$, in addition to the standard potential $A_{\mu}$ \cite{Cabibbo:1962td} (see also \cite{Zwanziger:1970hk,Barker:1978ck,Singleton:1996hgp}). While this alternative formulation of classical electrodynamics is not entirely necessary for understanding the double copy of type D spacetimes, it is nonetheless useful, particularly for the single copy interpretation of the charged C-metric. 

The role of the potential $C_{\mu}$ is to modify the Maxwell equations such that they are symmetric in electric and magnetic fields in the presence of magnetic charges and currents. Doing so, however, requires an extension of the field strength tensor $F_{\mu\nu}$ to incorporate both four-vector potentials  $A_\mu$ and $C_\mu$,
\begin{equation}
    F_{\mu\nu}=\partial_{[\mu}A_{\nu]}+\varepsilon_{\mu\nu}^{ \ \ \rho\sigma}\partial_\rho C_\sigma.
\end{equation}
This field strength and its dual satisfy 
\begin{equation}\label{modMax}
    \nabla_\nu F^{\mu\nu}=J^\mu_e, \ \ \ \ \ \nabla_\nu\tilde{F}^{\mu\nu}= J^\mu_m,
\end{equation}
as desired, where the new potential $C_{\mu}$ is solely responsible for the magnetic current $J_{m}^{\mu}$. The components of (\ref{modMax}) are
\begin{equation}
\begin{split} 
    \nabla\cdot\vec{E}=\rho_e, \ \ \ & \ \ \ \ \ \  \nabla\times\vec{B}=\vec{J}_{e}+\partial_t\vec{E}, \\
    \nabla\cdot\vec{B}=\rho_m, \ \ \ & \ \ \ -\nabla\times\vec{E}=\vec{J}_{m}+\partial_t\vec{B}. \\
    \end{split} 
\end{equation}

Let us now apply this two-potential formalism to the charged C-metric, where we can explicitly construct the gauge potentials $A_{\mu}$ and $C_{\mu}$ required to produce the field equations (\ref{dF}) and (\ref{dFtilde}).  The full field strength tensor $F^{(2)}_{\mu\nu}$ is
\begin{equation}
    F^{(2)}_{\mu\nu}=\partial_{[\mu}A^{(2)}_{\nu]}+\sqrt{-g^{(0)}}g_{(0)}^{\rho\lambda}g_{(0)}^{\sigma\gamma}\epsilon_{\mu\nu\lambda\gamma}\partial_\rho C_\sigma,
\end{equation}
with $g^{(0)}_{\mu\nu}$ being the accelerated Minkowski metric (\ref{AccelMink}).
Using that $F_{\mu\nu}^{(2)}$ takes the form in (\ref{eq:F2Cmet}),
we find that the two vector potentials are
\begin{equation}\label{twopotentialsforC}
    A^{(2)}_\mu = -\frac{q_2}{4r^2}\delta^t_\mu, \ \ \ \ \ \ C_\mu = -\frac{q_2\alpha}{4}\sin^2\theta\delta^\phi_\mu.
\end{equation}
 The $A^{(2)}_\mu$ leads to a correction to the Li\'enard-Wiechert potential (\ref{A1chargedC})\footnote{The Li\'enard-Wiechert potential in the accelerated coordinates falls off as $1/r$, while $A^{(2)}$ falls off as $1/r^2.$ Transforming out of the accelerated coordinates into standard Minkowski space in spherical coordinates $(T,R,\Theta,\Phi)$ centered on the point charge, it follows that the sum of $A^{(1)}$ and $A^{(2)}$ can be written as 
 $$A^{(1)}+A^{(2)}=\big[1+\frac{q_2}{q_1}\big(\alpha+\frac{2\cos^2\Theta}{\alpha R^2}+O(R^{-4})\big)\big]A^{\text{(LW)}},$$ 
 where $A^{\text{(LW)}}$ is the Li\'enard-Wiechert potential. Note that the $\alpha\rightarrow 0$ is not well defined because the diffeomorphism between the accelerated coordinates to standard Minkowski space is singular in that limit \cite{Luna:2018dpt}.}, while the $C_\mu$ accounts for the magnetic monopole-like behavior. Moreover, we obtain the Reissner-Nordstr\"om solution from the charged C-metric in the limit that $\alpha\rightarrow 0.$ Clearly, $C_\mu$ in (\ref{twopotentialsforC}) vanishes in that limit,
 and we recover traditional Maxwell theory.

\vspace{.5cm}

\noindent \textbf{Ricci Spinor}

From the zeroth and single copy fields (\ref{CmetWeyl}), (\ref{eq:S2Cmet}), and (\ref{eq:f2Cmet}), it is clear
\beq \frac{1}{S_{(1)}}f_{AB}^{(1)}\propto \frac{1}{S_{(2)}}f_{AB}^{(2)}\;,\quad S_{(2)}\not\propto|S_{1}|^{2}. \label{eq:CmetS2notS1}\eeq
As expected, the Ricci spinor (\ref{RicSpin}) is proportional to the modulus of the single copy spinors $f_{AB}^{(1)}$,
\begin{equation}
    f^{(1)}_{AB}\bar{f}^{(1)}_{C'D'}=q_1^2\frac{(1-\alpha r\cos\theta)^4}{r^4}o_{(A}\iota_{B)}\bar{o}_{(C'}\bar{\iota}_{D')}=\frac{q_{1}^{2}}{2Q^{2}} \Phi_{ABC'D'}.
\label{eq:cmetriccif1}\end{equation}
 However, 
\begin{equation}
    \frac{1}{S_{(2)}}f^{(2)}_{AB}\bar{f}^{(2)}_{C'D'}=6Q^{2}\frac{(1+\alpha r\cos\theta)(1-\alpha r\cos\theta)^3}{r^4}o_{(A}\iota_{B)}\bar{o}_{(C'}\bar{\iota}_{D')}\not\propto \Phi_{ABC'D'},
\label{eq:CmetRiccispinorf2f2}\end{equation}
as eluded to above, although the two expressions agree in the $\alpha\to0$ limit.



\section{General electro-vacuum type D spacetimes}\label{sec:gentypeD}


The most general Petrov type D solution of the Einstein-Maxwell field equations with a non-zero cosmological constant $\Lambda$ is described by the Plebanski-Demianski (PD) family of metrics \cite{Plebanski:1976gy}. A useful form of the line element\footnote{We use the form of the metric presented in \cite{Griffiths:2005qp} with a `mostly plus' signature. } is
\beq ds^{2}=\frac{1}{(1-pq)^{2}}\left[\frac{p^{2}+q^{2}}{P(p)}dp^{2}+\frac{p^{2}+q^{2}}{Q(q)}dq^{2}+\frac{P(p)}{p^{2}+q^{2}}(d\tau+q^{2}d\sigma)^{2}-\frac{Q(q)}{p^{2}+q^{2}}(d\tau-p^{2}d\sigma)^{2}\right]\;,\label{eq:PDmetorig}\eeq
with
\begin{equation}
\begin{split} 
    Q(q)&=k+e^2+g^2 - 2mq+\epsilon q^2-2nq^3 -\big(k+\Lambda /3\big)q^4\;,\\
    P(p)&=k+2np-\epsilon p^2+2mp^3-\big(e^2+g^2+k+\Lambda /3\big)p^4.
    \end{split}
\label{eq:formQPorig}\end{equation}
The real parameters $\{m,n,e,g,\epsilon,\Lambda\}$ are taken to be arbitrary, however, limiting cases of the PD metric, \emph{e.g.}, the Kerr-Newman black hole  provide a physical interpretation for these parameters; \emph{e.g.}, $m$ and $n$ correspond to mass and NUT parameters, respectively, and $e$ and $g$ correspond to electric and magnetic monopole charges. While the classical Kerr-Schild double copy has been extended to describe maximally symmetric spacetimes \cite{Carrillo-Gonzalez:2017iyj}, we will set $\Lambda=0$. For $e=g=\Lambda=0$ and $k=\gamma$ one recovers the PD metric analyzed in the pure Weyl double copy \cite{Luna:2018dpt}. 

Notably, the PD metric cannot generally be placed in Kerr-Schild form (\ref{KSmet}), however, it may be put into double Kerr-Schild form via the complex diffeomorphism \cite{Plebanski:1976gy}
\beq \tau=u+\int \frac{q^{2}dq}{Q(q)}+i\int\frac{p^{2}dp}{P(p)}\;,\quad \sigma=v-\int\frac{dq}{Q(q)}+i\int\frac{dp}{P(p)}\;,\eeq
such that the  metric (\ref{eq:PDmetorig}) takes the form
\beq ds^{2}=\frac{1}{(1-pq)^{2}}\left[2(iKdp-Ldq)+\frac{P(p)}{p^{2}+q^{2}}K^{2}-\frac{Q(q)}{p^{2}+q^{2}}L^{2}\right]\;.
\label{eq:complexpdmetv2}\eeq
where one has introduced null, geodesic, and mutually orthogonal covectors $K$ and $L$
\beq K=du+q^{2}dv\;, \quad  L=du-p^{2}dv\;.\eeq

The flat space limit is identified by simultaneously setting all dynamical, curvature producing parameters to zero, namely, $m=n=g=e=0$.
The remaining parameters, $k$ and $\epsilon$, are said to be \emph{kinematical} as they do not generate curvature. Following \cite{Luna:2018dpt}, we include the kinematical parameters in the definition of the flat space metric $ds^{2}_{(0)}$ which is identified as
\beq ds^{2}_{(0)}=\frac{1}{(1-pq)^{2}}\left[2(iKdp-Ldq)+\frac{k(1-p^{4})-\epsilon p^{2}}{(p^{2}+q^{2})}K^{2}-\frac{k(1-q^{4})+\epsilon q^{2}}{(p^{2}+q^{2})}L^{2}\right]\;.\label{eq:flatlimPD}\eeq
From here it is easy to see the PD metric in these coordinates is of double Kerr-Schild form with  Kerr-Schild functions $\phi_{K}$ and $\phi_{L}$
\beq \phi_{K}=\frac{2np+2mp^{3}-(e^{2}+g^{2})p^{4}}{(p^{2}+q^{2})(1-pq)^{2}}\;,\;\; \phi_{L}=\frac{2mq+2nq^{3}-e^{2}-g^{2}}{(p^{2}+q^{2})(1-pq)^{2}}\;,\label{eq:KerrschildfuncsPD}\eeq
and hence
\beq ds^{2}=ds^{2}_{(0)}+\phi_{K}K^{2}+\phi_{L}L^{2}\;.\eeq
Keeping the kinematical parameters in the flat space portion of the double Kerr-Schild decomposition is a choice, and thus far, ambiguous. We will return to this ambiguity in a moment and see how it affects the source contributions to the Weyl double copy.

While the Ricci tensor $R^{\mu}_{\;\nu}$ is linear in Kerr-Schild form, this property is generally not the case for metrics in double Kerr-Schild form. Nonetheless, as shown in \cite{Luna:2018dpt}, the PD metric (\ref{eq:complexpdmetv2}) has a linear Ricci tensor, given by
\begin{equation}\label{mixedRicci}
R^\mu_{ \ \nu}=(e^2+g^2)\left(
\begin{array}{cccc}
 \frac{(p q-1)^4 \left(p^2-q^2\right)}{\left(p^2+q^2\right)^3} & \frac{2 p^2 q^2  (p q-1)^4}{\left(p^2+q^2\right)^3} & 0 & 0 \\
 \frac{2  (p q-1)^4}{\left(p^2+q^2\right)^3} & -\frac{ (p q-1)^4 \left(p^2-q^2\right)}{\left(p^2+q^2\right)^3} & 0 & 0 \\
 0 & 0 & \frac{ (p q-1)^4}{\left(p^2+q^2\right)^2} & 0 \\
 0 & 0 & 0 & -\frac{(p q-1)^4}{\left(p^2+q^2\right)^2} \\
\end{array}
\right)
\end{equation}
from which we obtain the energy density $\rho^{\text{grav}}$, \emph{i.e.}, the negative of the first entry of (\ref{mixedRicci}),
\begin{equation}\label{rhograv}
    \rho^{\text{grav}}=  \frac{\left(e^2+g^2\right) (p q-1)^4 \left(q^2-p^2\right)}{\left(p^2+q^2\right)^3}  .
\end{equation}
With the Ricci tensor we can build its spinorial counterpart \cite{Griffiths:2005qp},
\begin{equation}\label{PDriccispin} \Phi_{ABC'D'}=2(e^{2}+g^{2})\frac{(1-pq)^{4}}{(q^{2}+p^{2})^{2}}o_{(A}\iota_{B)}\bar{o}_{(C'}\bar{\iota}_{D')}.
\end{equation} 
Lastly, the Weyl spinor corresponding to the metric (\ref{eq:complexpdmetv2}) is \cite{Griffiths:2005qp}
\begin{equation}\label{Psi2PD}
\begin{split} 
    \Psi_{ABCD}& =\Psi^{(1)}_{ABCD}+\Psi^{(2)}_{ABCD} \\
    &=-6(m+in)\Bigg(\frac{1-pq}{q+ip}\Bigg)^3o_{(A}o_B\iota_C\iota_{D)}+6(e^2+g^2)\Bigg(\frac{1-pq}{q+ip}\Bigg)^3\frac{1+pq}{q-ip}o_{(A}o_B\iota_C\iota_{D)}.
    \end{split} 
\end{equation}
The first term $\Psi^{(1)}_{ABCD}=\frac{1}{S_{(1)}}f^{(1)}_{(AB}f^{(1)}_{CD)}$ was previously determined in \cite{Luna:2018dpt}, explicitly given by
\begin{equation}
    S_{(1)}=\frac{i}{6}\frac{(\tilde{m}+i\tilde{n})^2(1-pq)}{(m+in)(p-iq)}, \ \ \ \ \ f^{(1)}_{AB}=\frac{(\tilde{m}+i\tilde{n})(1-pq)^2}{(p-iq)^2}o_{(A}\iota_{B)}\;,
\label{eq:gaugefields1PD}\end{equation}
 and satisfy the vacuum equations over the flat background (\ref{eq:flatlimPD}). The constants $\tilde{m}$ and $\tilde{n}$ are introduced in the spirit of having gauge parameters separate from the gravity parameters.

\vspace{.5cm}


\noindent \textbf{Sourced Weyl double copy} 

We analyzed the second term (\ref{Psi2PD}) in \cite{Easson:2021asd}:
\begin{equation}
    S_{(2)}=\frac{(\tilde{e}^2+\tilde{g}^2)^2}{6(e^2+g^2)} \Bigg(\frac{1-pq}{q+ip}\Bigg)\frac{1+pq}{q-ip}, \ \ \ \ \ \ \ f^{(2)}_{AB}=(\tilde{e}^2+\tilde{g}^2)\Bigg( \frac{1-pq}{q+ip}\Bigg)^2\frac{1+pq}{q-ip}o_{(A}\iota_{B)},
\label{eq:gaugefields2PD}\end{equation}
introducing gauge parameters $\tilde{e}$ and $\tilde{g}$. Notice the wave equation of the zeroth copy scalar field obeys
\begin{equation}
\begin{split} 
    \Box^{(0)}S_{(2)}&=\frac{(\tilde{e}^2+\tilde{g}^2)^2}{6(e^2+g^2)}\Bigg(-2\epsilon\frac{(1-pq)^4(q^2-p^2)}{(p^2+q^2)^3}+4k\frac{(1-pq)^4(1+p^2q^2)}{(p^2+q^2)^3} \Bigg)\\
    &\propto \epsilon\rho^{\text{grav}}+k\Delta (p,q).
    \end{split}
\end{equation}
This shows $\Box^{(0)}S_{(2)}\propto\rho^{\text{grav}}$ is met \textit{only} if the kinematical parameter vanishes, $k\rightarrow 0.$ Due to the ambiguity in choosing the flat background, this is no issue. Indeed, the vacuum equations for $S_{(1)}$ and $f^{(1)}_{AB}$ are immune to the choice of $k$. This observation suggests the sourced Weyl double copy may specify an appropriate flat background where the single and zeroth copy fields live. 


The Maxwell tensor associated to the spinor field strength $f^{(2)}_{AB}$ is
\begin{equation}\label{uvF}
    F_{\mu\nu}^{(2)}=(\tilde{e}^2+\tilde{g}^2)
\left(
\begin{array}{cccc}
 0 & 0 & \frac{p (p q+1)}{2 \left(p^2+q^2\right)^2} & \frac{q (p q+1)}{2 \left(p^2+q^2\right)^2} \\
 0 & 0 & \frac{p q^2 (p q+1)}{2 \left(p^2+q^2\right)^2} & -\frac{p^2 q (p q+1)}{2 \left(p^2+q^2\right)^2} \\
 -\frac{p (p q+1)}{2 \left(p^2+q^2\right)^2} & -\frac{p q^2 (p q+1)}{2 \left(p^2+q^2\right)^2} & 0 & 0 \\
 -\frac{q (p q+1)}{2 \left(p^2+q^2\right)^2} & \frac{p^2 q (p q+1)}{2 \left(p^2+q^2\right)^2} & 0 & 0 \\
\end{array}
\right)
\end{equation}
and obeys
\begin{equation}\label{uvMax}
    \nabla^{(0)}_\mu F_{(2)}^{\nu\mu}=(\tilde{e}^2+\tilde{g}^2)\begin{pmatrix} -\frac{(q^2-p^2) (p q-1)^4}{2 \left(p^2+q^2\right)^3} \\ \frac{(p q-1)^4}{\left(p^2+q^2\right)^3} \\
    0 \\
    0
    \end{pmatrix}.
\end{equation}
Thus, (\ref{uvMax}) shows the single copy charge density is proportional to the gravitational energy density, and the current density is proportional to an angular momentum term, \emph{i.e.}, $R^{\mu}_{\;0}\propto J^{\mu}_{(2)}$,
as anticipated from the Kerr-Schild relation (\ref{sourceMaxInKS}). We also find that the Jacobi identity for (\ref{uvF}) fails, $\nabla^{(0)}_{[\mu}F^{(2)}_{\lambda\gamma]}\neq 0.$ Or, in terms of the divergence of $\tilde{F}^{(2)}_{\mu\nu}$,
\begin{equation}\label{divFdual}
    \nabla^{(0)}_{\nu}\tilde{F}_{(2)}^{\mu\nu} =(\tilde{e}^2+\tilde{g}^2)\begin{pmatrix}-\frac{p^2 q^2 (p q-1)^4}{\left(p^2+q^2\right)^3} \\
    \frac{(p^2-q^2) (p q-1)^4}{2 \left(p^2+q^2\right)^3} \\
    0 \\ 0 
    \end{pmatrix}\;.
\end{equation}
A non-zero divergence of $\tilde{F}^{\mu\nu}_{(2)}$ suggests the presence of a magnetic charge and magnetic current. This is consistent with what we found for the charged C-metric, which is a limiting case of the general PD metric. That is, since the PD metric is not asymptotically flat, the single copy gauge fields are sourced by both an electric and magnetic current density.\footnote{In \cite{Easson:2021asd} we imprecisely remarked the appearance of the magnetic monopole arises due to the moving NUT charge. Now it is clear the reason for the magnetic monopole-like behavior is primarily due to the acceleration.}

\vspace{.5cm}

\noindent \textbf{Ricci spinor}

 Using single copy spinor field strength $f^{(1)}_{AB}$ (\ref{eq:gaugefields1PD}), we have
\beq f^{(1)}_{AB}\bar{f}^{(1)}_{C'D'}=\frac{(\tilde{m}^{2}+\tilde{n}^{2})}{2(e^{2}+g^{2})}\Phi_{ABC'D'}, \eeq
which is clear by inspection to (\ref{PDriccispin}). Calling $(\tilde{m}^{2}+\tilde{n}^{2})\equiv q_{1}^{2}$ and $(e^{2}+g^{2})\equiv Q^{2}$, we uncover the same relation as in (\ref{eq:kerrnewriccif1}) and (\ref{eq:cmetriccif1}). Meanwhile, the non-vacuum single copy fields (\ref{eq:gaugefields2PD}) are not directly proportional to the Ricci spinor. This is consistent with the our Ricci spinor analysis of the charged C-metric (\ref{eq:CmetRiccispinorf2f2}), and is a byproduct of the fact $S_{(2)}\not\propto|S_{(1)}|^{2}$ in the general PD metric.

\subsection{Kerr-Newman-Taub-NUT}\label{sec:KNTN}

As our final example, let us consider a the Kerr-Newman-Taub-NUT solution, a special limiting case of the general PD metric. One can acquire the line element of this solution via a combination of coordinate and parameter rescalings of the general PD metric (\ref{eq:complexpdmetv2}). Specifically, perform the coordinate rescalings 
\beq u\to \ell u\;, \;v\to \ell^{3} v\;, \;p\to \ell^{-1}p\;,\; q\to \ell^{-1} q\;,\label{eq:coordrescalTN}\eeq
along with the parameter rescalings
\beq m\to \ell^{-3}m\;, \;n\to \ell^{-3}n\;,\; e\to \ell^{-2}e\;,\; g\to \ell^{-2}g\;,\;\epsilon\to\ell^{-2}\epsilon\;, \; k\to \ell^{-4}k\;.\label{eq:paramrescalTN}\eeq
Then, taking the $\ell\to\infty$ limit yields the line element
\beq
\begin{split}
 ds^{2}&=2(iKdp-Ldq)+k(2du+dv(q^{2}-p^{2}))dv-\epsilon(du^{2}+p^{2}q^{2}dv^{2})+\phi_{K}K^{2}+\phi_{L}L^{2}\;,
\end{split}
\label{eq:KNTNmet1} \eeq
where the first three terms correspond to the flat space line element and now the Kerr-Schild functions (\ref{eq:KerrschildfuncsPD}) $\phi_{K}$ and $\phi_{L}$ are
\beq \phi_{K}=\frac{2np}{p^{2}+q^{2}}\;,\quad \phi_{L}=\frac{2mq-(e^{2}+g^{2})}{p^{2}+q^{2}}\;.\eeq
Technically, the line element (\ref{eq:KNTNmet1}) is the Kerr-Newman-Taub-NUT solution, including rotation, electric, magnetic, and NUT charges (when $e=g=0$ we recover the form of the neutral Kerr-Taub-NUT solution considered in \cite{Luna:2018dpt}). Additionally, while this metric does not approach Minkowski space in the asymptotic limit, it nonetheless has vanishing Riemann curvature at spatial infinity.\footnote{This is mostly easily verified by switching to Boyer-Lindquist form coordinates; see, \emph{e.g.} \cite{Griffiths:2005qp}.}

Analogous to \cite{Luna:2018dpt}, the zeroth and single copy fields for the Kerr-Newman-Taub-NUT solution readily follow from applying the rescalings (\ref{eq:coordrescalTN}) and (\ref{eq:paramrescalTN}), together with rescaling $(\tilde{e},\tilde{g})\to \ell^{-2}(\tilde{e},\tilde{g})$, to the gauge fields of the general PD solution (\ref{eq:gaugefields1PD}) and (\ref{eq:gaugefields2PD}). Precisely, upon rescaling $S_{(1)}$ and $f_{AB}^{(1)}$, take the $\mathcal{O}(\ell^{-2})$ and $\mathcal{O}(\ell^{-1})$ coefficients, respectively, in a large $\ell$ limit,
\beq S_{(1)}=\frac{i}{6}\frac{(\tilde{m}+i\tilde{n})^{2}}{(m+in)(p-iq)}\;,\quad f_{AB}^{(1)}=\frac{\tilde{m}+i\tilde{n}}{(p-iq)^{2}}o_{(A}\iota_{B)}\;.\eeq
Likewise, we find
\beq S_{(2)}=\frac{(\tilde{e}^{2}+\tilde{g}^{2})^{2}}{6(e^{2}+g^{2})(p^{2}+q^{2})}\;,\quad f^{(2)}_{AB}=-i\frac{(\tilde{e}^{2}+\tilde{g}^{2})}{(p-iq)^{2}(p+iq)}o_{(A}\iota_{B)}\;.\eeq
It is worth pointing out that for real parameters $e,g,\tilde{e}$ and $\tilde{g}$, the function $S_{(2)}$ is a real function, unlike the zeroth copy $S_{(1)}$. Additionally,
\begin{equation}
    \frac{1}{S_{(1)}}f^{(1)}_{AB}\propto \frac{1}{S_{(2)}}f^{(2)}_{AB}\;,\quad  S_{(2)}\propto |S_{(1)}|^2\;,
\label{eq:S2S1eqKNTN}\end{equation}
unlike the relation (\ref{eq:CmetS2notS1}) for the charged C-metric and general PD solution,

In the same vein, upon the rescalings (\ref{eq:coordrescalTN}) and (\ref{eq:paramrescalTN}) and taking the $\ell\to\infty$ limit of (\ref{PDriccispin}) and (\ref{Psi2PD}), the Ricci spinor and Weyl spinor are, respectively, 
\beq \Phi_{ABC'D'}=2\frac{(e^{2}+g^{2})}{(p^{2}+q^{2})^{2}}o_{(A}\iota_{B)}\bar{o}_{(C'}\bar{\iota}_{D')}.\label{eq:RicciSpinKNTN}\eeq

\beq \Psi_{ABCD}=-6i\frac{(m+in)}{(p-iq)^{3}}o_{(A}o_B\iota_C\iota_{D)}-\frac{6(e^{2}+g^{2})}{(p-iq)^{3}(p+iq)}o_{(A}o_B\iota_C\iota_{D)},\label{eq:Psi2KNTN}\eeq
And, by virtue of (\ref{eq:S2S1eqKNTN}), we see
\beq \frac{1}{S_{(2)}}f^{(2)}_{AB}\bar{f}^{(2)}_{C'D'}=\frac{6(e^{2}+g^{2})}{(p^{2}+q^{2})^{2}}o_{(A}\iota_{B)}\bar{o}_{(C'}\bar{\iota}_{D')}=3\Phi_{ABC'D'},\eeq
which is the same form as the Kerr-Newman solution (\ref{RicciDC}) and is proportional to $f_{AB}^{(1)}\bar{f}^{(1)}_{C'D'}$.

\section{Discussion} \label{sec:disc}

In this work, we explored the sourced Weyl double copy \cite{Easson:2021asd} for exact solutions to Einstein-Maxwell theory. In doing so we provided the first double copy analysis of the charged C-metric and Kerr-Newman-Taub-NUT black hole, and developed a double prescription for the Ricci spinor for the type D solutions considered. Paramount in our interpretation of the zeroth and single copy fields was the role played by Riemann asymptotic flatness, \emph{i.e.}, whether the Riemann curvature tensor was non-vanishing at spatial infinity. Spacetimes whose Riemann curvature tensor was non-vanishing asymptotically, namely, the charged C-metric and general Plebanski-Demianski solution, had a corresponding single copy dual field strength tensor with non-vanishing divergence. This indicates the existence of a non-trivial magnetic current. In the case of the charged C-metric, we used a two-potential formulation of electrodynamics to provide a consistent description of the single copy field strength, $F^{(2)}_{\mu\nu}$, leading to a correction of the Li\'enard-Wiechert potential (owed to $A_{\mu}^{(2)}$) and a second potential $C_\mu$ responsible for the magnetic current density. Additionally,  we showed all Riemann asymptotically flat electrovacuum spacetimes, including the Kerr-Newman-Taub-NUT solution, have a Ricci spinor whose double copy form is reminiscent of the Weyl double copy. Summarily, 
\beq \Phi_{ABC'D'}=\frac{2 q^{2}_{s}}{q_{1}^{2}} f^{(1)}_{AB}\bar{f}^{(1)}_{C'D'}\;,\quad \Phi_{ABC'D'}=\frac{1}{3S_{(2)}}f^{(2)}_{AB}\bar{f}^{(2)}_{C'D'},\eeq
where the first equality holds for all electrovacuum solutions with external source charge $q_{s}$, while the second equality only holds for asymptotically flat electrovacuum spacetimes.

There are multiple potential avenues of future research. Firstly, it would be worthwhile to examine the sourced Weyl double copy including matter field sources beyond Einstein-Maxwell theory. More generally one may  include additional gauge fields, scalar fields, etc., motivated by, \emph{e.g.}, string theory or supergravity. Such extensions have been considered in the context of the perturbative double copy, however, the exact Kerr-Schild double copy for pure Einstein gravity has difficulty describing an additional massless NS-NS field like a Kalb-Ramond field or a dilaton using the single null congruence in the conventional Kerr-Schild ansatz. To understand the double copy of such fields in a non-perturbative way, the Kerr-Schild framework was generalized to describe double-field theory and supergravities in \cite{Lee:2018gxc,Cho:2019ype}. It  would be interesting to try to incorporate such fields using the sourced Weyl double copy, and compare to the spinorial representation of the torsionful Riemann curvature introduced in \cite{Monteiro:2021ztt} to study solutions of the universal massless sector of supergravity. Doing so may also lead to a better understanding how the sourced Weyl double copy relates to the scattering amplitudes of the corresponding single copy gauge field theory. 


Further, it would be interesting to extend our current analysis to other types of black hole spacetimes, particularly those including a cosmological constant (whose Weyl double copy  was presented in \cite{Han:2022mze}), additional Abelian gauge fields \cite{Chawla:2022ogv}, and  massless or higher spin fields \cite{Han:2022ubu,Didenko:2022qxq}. It would also be interesting to adapt our formalism to three-dimensional spacetimes, such as the BTZ black hole \cite{Banados:1992wn}, its quantum and de Sitter generalizations \cite{Emparan:2020znc,Emparan:2022ijy}, or generalized three-dimensional solutions described in \cite{Alkac:2022tvc,Bueno:2021krl}. To this end, it would be worthwhile to also attempt to extend the Cotton double copy \cite{Emond:2022uaf} to include external sources.

Finally, we have studied an elementary extension of the original Weyl double copy, and begun to gather evidence of a double copy prescription for the Ricci spinor. To better understand both double copy  structures on a more fundamental level, it would be interesting to explore the twistorial foundations of the sourced Weyl double copy, utilizing and extending \cite{White:2020sfn,Chacon:2021wbr,Chacon:2021hfe,Chacon:2021lox,Guevara:2021yud,Farnsworth:2021wvs,Luna:2022dxo}.

\vspace{1cm}

\noindent {\large \textbf{Acknowledgements:}} We are grateful to Gabriel Herczeg for useful discussion while this work was in progress. DE is supported in part by the U.S. Department of Energy, Office of High Energy Physics, under Award No. DE-SC0019470, and the Foundational Questions Institute under Grant number FQXi-MGB-1927. TM is supported by the Simons Foundation, Award 896696. AS is supported by the Simons Foundation via the \emph{It from Qubit collaboration on quantum fields, gravity, and information} and EPSRC.

\appendix

\section{Spinor formalism}\label{app:Aspinorformalism}

Here we briefly illustrate how to obtain various spinorial quantities of interest from a four-dimensional curved spacetime metric $g_{\mu\nu}$. In our notation, spacetime indices are given by $\{\mu,\nu,...\}$, frame indices by $\{a,b,...\}$, while the spinor indices are $\{A,B,...\}$ and their conjugates $\{A',B',...\}$. Our remaining conventions primarily follow \cite{Stephani:2003tm,Penrose:1985bww}.

\begin{center}
    \textbf{Tetrads and the spinor basis}
\end{center}

We introduce a complex null tetrad $\{l,n,m,\bar{m}\}$ constructing the metric $g_{\mu\nu}$ and satisfies
\begin{equation}\label{NTrequirements}
    \begin{split} 
g_{\mu\nu}&=-2l_{(\mu}n_{\nu)}+2m_{(\mu}\bar{m}_{\nu)}, \\
l_{\mu}l^{\mu}&=n_{\mu}n^{\mu}=m_{\mu}m^{\mu}=\brm_{\mu}\brm^{\mu}=0,\\
n^{\mu}l_{\mu}&=-1, \ \ \ \ \ \ \ \ \ m^{\mu}\brm_{\mu}=+1,
\end{split}
\end{equation}
 for which null vectors $n^{\mu}$ and $l^{\mu}$ are real and $m^{\mu}$ is generally complex with $\bar{m}^{\mu}$ its conjugate. The tetrad vectors have an associated frame tetrad that can be obtained using the vierbein as, \textit{e.g.}, $l^a=e^a_{ \ \mu}l^\mu.$ Explicitly, we work with the tetrad set
 \begin{equation}\label{frametetrads}
     \begin{split}
         l_a=\frac{1}{\sqrt{2}}(1,-1,0,0), \ \ & \ \  n_a=\frac{1}{\sqrt{2}}(1,1,0,0) \\
         m_a=\frac{1}{\sqrt{2}}(0,0,i,1), \ \ & \ \ \bar{m}_a=\frac{1}{\sqrt{2}}(0,0,-i,1),
     \end{split}
 \end{equation}
 which constructs Minkowski space analogously to (\ref{NTrequirements}) as $\eta_{ab}=-2l_{(a}n_{b)}+2m_{(a}\bar{m}_{b)}$. All frame indices are raised with $\eta^{ab}$.
 
 To translate between tensors and spinors, we use the Pauli four-vectors 
 \begin{equation}\label{Pauli4}
     \s^a_{AA'}=\frac{1}{\sqrt{2}}(1,\vec{\s})_{AA'}, 
 \end{equation}
 where the $\s_i$ are the standard $SU(2)$ generators 
\begin{equation} 
\sigma_{1}\doteq \begin{pmatrix}0 & 1 \\ 1 & 0 \end{pmatrix}, \ \ \ \sigma_{2} \doteq\begin{pmatrix}0 & -i \\ i & 0 \end{pmatrix}, \ \ \ \sigma_{3}\doteq\begin{pmatrix}1 & 0 \\ 0 & -1 \end{pmatrix}\;.
\end{equation}
The Pauli four-vectors satisfy
\begin{equation}
    \s^a_{AA'}\s_a^{BB'}=\delta^A_B\delta^{A'}_{B'}, \ \ \ \ \s^a_{AA'}\s_b^{BB'}=\delta^a_b.
\end{equation}
Then, any spacetime (frame) vector has a spinor analog, 
\begin{equation}
    V_a \ \rightarrow \ V_{AA'}=V_a \s^a_{AA'},
\end{equation}
which has its associated spacetime vector $V_\mu=e_\mu^{a}V_a.$ 

We next identify a spinor basis $\{o_A,\iota_A\}$ (and the conjugate basis $\{\bar{o}_{A'}\bar{\iota}_{A'}\}$), whose indices are raised and lowered by the two-dimensional Levi-Civita symbol
\begin{equation}
    \epsilon^{AB}\doteq\begin{pmatrix}0 & 1 \\ -1 & 0 \end{pmatrix}=-\epsilon_{AB} 
\end{equation}
and its conjugate $\epsilon^{A'B'}$. The basis spinors are related to the frame tetrad by 
\begin{equation}\label{tetsandspins}
\begin{split} 
    o_A\bar{o}_{A'}=l_a\s^a_{AA'}, \ \ & \ \ \iota_A\bar{\iota}_{A'}=n_a\s^a_{AA'}, \\
    \iota_A\bar{o}_{A'}=m_a\s^a_{AA'}, \ \ & \ \ o_A\bar{\iota}_{A'}=\bar{m}_{a}\s^a_{AA'}.
    \end{split}
\end{equation}
The spinor basis satisfies $\epsilon^{AB}o_Bo_A=\epsilon^{AB}\iota_B\io_A=0,$ $\epsilon^{AB}\io_Bo_A=\io^Ao_A=1,$ with $\epsilon^{AB}o_B\io_A=o^A\io_A=-1$. Using (\ref{tetsandspins}) along with our choice of tetrads (\ref{frametetrads}), we can deduce that the (normalized) spinor basis vectors are given by
\begin{equation}
    o_A=\frac{1}{\sqrt{2}}(1,1), \ \ \ \ \io_A=\frac{1}{\sqrt{2}}(1,-1).
\end{equation}

We next introduce the Infeld-van der Waerden symbols
\begin{equation}\label{IvdWsymbols}
    \s^{ab}_{AB}=\s^{[a}_{AA'}\bar{\s}^{b]A'C}\epsilon_{CB}, \ \ \ \ \ \ \bar{\s}=\frac{1}{\sqrt{2}}(1,-\vec{\s}),
\end{equation}
which, along with the spacetime vierbeins, allow us to obtain the spinorial forms of any (even) rank-2 and higher tensor. For example, defining $\s^{\mu\nu}_{AB}=e^\mu_ae^\nu_b\s^{ab}_{AB},$ the Weyl spinor and spinor field strength are given by
\begin{equation}
    \Psi_{ABCD}=\frac{1}{4}W_{\mu\nu\alpha\beta}\s^{\mu\nu}_{AB}\s^{\alpha\beta}_{CD},
\end{equation}
\begin{equation}\label{fAB}
    f_{AB}=\frac{1}{2}F_{\mu\nu}\s^{\mu\nu}_{AB},
\end{equation}
where $W_{\mu\nu\alpha\beta}$ is the Weyl tensor and $F_{\mu\nu}$ is the standard field strength tensor. Both $\Psi_{ABCD}$ and $f_{AB}$ are completely symmetric in their indices.


\vspace{.5cm}


\begin{center}
    \textbf{Field strength spinor}
\end{center}

In the Weyl double copy, the spinor field strength lives in an appropriate flat limit of the full spacetime metric, which we relate to the `Minkowski vierbeins' $e^{(0)}$ as $g_{\mu\nu}^{(0)}=e_\mu^{(0),a}e_\nu^{(0),b}\eta_{ab}$. We then define the `frame field strength' $F_{ab}$, related to the spacetime field strength by
\begin{equation}
   F_{\mu\nu}=e_\mu^{(0),a}e_\nu^{(0),b}F_{ab}.
\end{equation}
Importantly, $F_{ab}$ lives on Minkowski space defined by $\eta_{ab}=\text{diag}(-1,1,1,1)$, while $F_{\mu\nu}$ lives on a different form of flat space described by $g^{(0)}_{\mu\nu}.$

The spinor field strength associated with the frame field strength $F_{ab}$ is analogous to (\ref{fAB}),
\begin{equation}
    f_{AB}=\frac{1}{2}F_{ab}\s^{ab}_{AB}.
\end{equation}
Component-wise, $f_{AB}$ is 
\begin{equation}\label{fABs}
    \begin{split}
        f_{00}&=-F_{01}+F_{13}+i(F_{02}-F_{23}), \\
        f_{01}&=F_{03}+iF_{12} \\
        f_{11}&=F_{01}+F_{13}+i(F_{02}+F_{23}),
    \end{split}
\end{equation}
We can invert (\ref{fABs}) together with their complex conjugate $\bar{f}_{A'B'}$ to solve for the $F_{ab}$’s:
\begin{equation}\label{Ftof}
    \begin{split}
   &F_{01}=\frac{1}{4}\Big[f_{11}-f_{00}+\bar{f}_{1'1'}-\bar{f}_{0'0'}\Big], \quad F_{02}=\frac{1}{4i}\Big[\bar{f}_{1'1'}+\bar{f}_{0'0'}-\Big(f_{11}+f_{00}\Big)\Big], \quad F_{03}=\frac{1}{2}\Big[f_{01}+\bar{f}_{0'1'}\Big] \\
&F_{13}=\frac{1}{4}\Big[f_{00}+f_{11}+\bar{f}_{0'0'}+\bar{f}_{1'1'}\Big],\quad  F_{23}=\frac{1}{4i}\Big[f_{11}-f_{00}-\Big(\bar{f}_{1'1'}-\bar{f}_{0'0'}\Big)\Big],\quad F_{12}=\frac{1}{2i}\Big[f_{01}-\bar{f}_{0'1'}\Big].
  \end{split}
\end{equation}
Note that for each of the type D metrics we consider, the matrix structure of $f_{AB}$ is the Pauli matrix $\s_3$, \textit{i.e.}, 
\begin{equation}
    f_{AB}=Z\begin{pmatrix}1&0\\0&-1\end{pmatrix}_{AB},
\end{equation}
where in general, $Z\in\mathbb{C}$. Thus, $f_{00}=-f_{11}=Z,$ $f_{0'0'}=-f_{1'1'}=Z^{*}$, and $f_{10}=f_{01}=0.$ This allows us to write the two nonzero components of $F_{ab}$ solely in terms of $Z$  as
\begin{equation}
    F_{01}=-\frac{1}{2}\text{Re}(Z), \ \ \ \ \ \ \ F_{23}=-\frac{1}{2}\text{Im}(Z).
\end{equation}

Finally, to obtain $F_{\mu\nu}$, we now contract $F_{ab}$ terms using the flat space vierbeins $e_\mu^{(0),a}$. Dropping the (0)-superscript on the vielbeins for conciseness, this reads
\begin{equation}\label{GenTypeDFmunu}
    F_{\mu\nu}=F_{ab}e_{\mu}^a e_\nu^b =-\frac{1}{2}\Big(\text{Re}(Z)e_{[\mu}^0 e_{\nu]}^1+\text{Im}(Z)e_{[\mu}^2 e_{\nu]}^3\Big).
\end{equation}
We point out that when vierbeins are diagonal (as would be the case if $g^{(0)}_{\mu\nu}$ is the metric in spherical polar or oblate spheroidal coordinates), then, from (\ref{GenTypeDFmunu}),  if $Z$ has an imaginary component, then $F_{\mu\nu}$ has a magnetic field component. This is in line with standard results for the Schwarzschild solution contrasted to the Kerr solution: the presence of the rotation parameter resulted in $Z$ being complex in the latter case, which corresponds to the magnetic field present as a result of the angular momentum of the black hole. For Schwarzschild, $Z\in\mathbb{R}$ and the single copy gauge field strength contains a (radial) electric field only.

Finally, as described in section \ref{sec:U(1)prescription}, let us explicitly show the form of the spinor field strengths of the single copy source $f^{(1)}_{AB}$ and double copy source $f^{s}_{AB}$ are equivalent. For frame field strength $F^{(1)}_{ab}=F^{(1)}_{\mu\nu}(e^{(0)})^{\mu}_{a}(e^{(0)})_{b}^{\nu}$, then
\begin{equation}\label{f1frameFapp}
    \begin{split}
        f^{(1)}_{AB}&=\frac{1}{2}F^{(1)}_{\mu\nu}(e^{(0)})^\mu_a(e^{(0)})^\nu_b\sigma^{[a}_{AC'}\bar{\sigma}^{b]C'}_{ \ \ B} \\
        &=\frac{1}{2}F^{(1)}_{cd}(e^{(0)})^{c}_\mu (e^{(0)})^{d}_\nu (e^{(0)})^\mu_a(e^{(0)})^\nu_b\sigma^{[a}_{AC'}\bar{\sigma}^{b]C'}_{ \ \ B} \\
        &=\frac{1}{2}F^{(1)}_{cd}\delta^c_a\delta^d_b\sigma^{[a}_{AC'}\bar{\sigma}^{b]C'}_{ \ \ B} \\
        &=\frac{1}{2}F^{(1)}_{ab}\sigma^{[a}_{AC'}\bar{\sigma}^{b]C'}_{ \ \ B},
    \end{split}
\end{equation}
where we used $(e^{(0)})_{\mu}^{a}(e^{(0)})^{\mu}_{b}=\delta^{a}_{b}$ in the third line. Likewise, \begin{equation}\label{fsframeFapp}
    \begin{split}
        f^{s}_{AB}&=\frac{1}{2}F^{s}_{\mu\nu}e^\mu_ae^\nu_b\sigma^{[a}_{AC'}\bar{\sigma}^{b]C'}_{ \ \ B} \\
        &=\frac{1}{2}F^{s}_{cd}e^{c}_\mu e^{d}_\nu e^\mu_ae^\nu_b\sigma^{[a}_{AC'}\bar{\sigma}^{b]C'}_{ \ \ B} \\
        &=\frac{1}{2}F^{s}_{cd}\delta^c_a\delta^d_b\sigma^{[a}_{AC'}\bar{\sigma}^{b]C'}_{ \ \ B} \\
        &=\frac{1}{2}F^{s}_{ab}\sigma^{[a}_{AC'}\bar{\sigma}^{b]C'}_{ \ \ B},
    \end{split}
\end{equation} 
where we used $e_{\mu}^{a}e^{\mu}_{b}=\delta^{a}_{b}$.



\begin{center}
    \textbf{Weyl spinor}
\end{center}

\vspace{.5cm}

 The  Weyl spinor $\Psi_{ABCD}$ may  be expanded using the spinor basis $\{o_A,\iota_B\}$ as \cite{Penrose:1985bww,Penrose:1986ca}
\begin{equation}\label{FullWeylSpinor}
    \Psi_{ABCD}=\Psi_0\iota_A\iota_B\iota_C\iota_D-4\Psi_1o_{(A}\iota_B\iota_C\iota_{C)}+6\Psi_2o_{(A}o_B\iota_C\iota_{D)}-4\Psi_3 o_{(A}o_Bo_C\iota_{D)}+\Psi_4 o_Ao_Bo_Co_D.
\end{equation}
Here parentheses represent symmeterization, with the convention, $H_{(AB)}=\frac{1}{2}(H_{AB}+H_{BA})$. The $\Psi_I\in\mathbb{C}$ denote the Weyl scalars involved in the Petrov classification of the spacetime (see, \textit{e.g.} \cite{Stephani:2003tm} for a thorough description of the classification scheme), and are defined using the Weyl tensor $W_{\mu\nu\lambda\gamma}$ and the complex null tetrad (\ref{NTrequirements})
\begin{equation}\label{PsiIs}
\begin{split}
    \Psi_{0}&=W_{\mu\nu\rho\lambda}n^{\mu}m^{\nu}n^{\rho}m^{\lambda} , \\
    \Psi_{1}&=W_{\mu\nu\rho\lambda}n^{\mu}l^{\nu}n^{\rho}m^{\lambda}, \\
    \Psi_{2}&=W_{\mu\nu\rho\lambda}n^{\mu}m^{\nu}\brm^{\rho}l^{\lambda}, \\
    \Psi_{3}&=W_{\mu\nu\rho\lambda}n^{\mu}l^{\nu}\brm^{\rho}l^{\lambda}, \\
    \Psi_{4}&=W_{\mu\nu\rho\lambda}\brm^{\mu}l^{\nu}\brm^{\rho}l^{\lambda}.
\end{split}    
\end{equation}
For all Petrov type D spacetimes, one can always choose a coordinate system such that $\Psi_2\neq 0$ while all other $\Psi_I$ vanish, such that the Weyl spinor becomes 
\begin{equation}
    \Psi_{ABCD}^{\text{type D}}=6\Psi_2o_{(A}o_B\io_C\io_{D)}.
\end{equation}



\vspace{.5cm}
\begin{center}
    \textbf{Ricci spinor}
\end{center}

The Riemann tensor decomposes into the Weyl tensor, Ricci tensor, and Ricci scalar schematically as $Riem = Weyl + Ric\otimes g+R(g\otimes g)$. Similarly, the Riemann spinor decomposes as
\begin{equation}\label{RiemannSpinor}
    \begin{split}
        R_{AA'BB'CC'DD'}&=\Psi_{ABCD}\epsilon_{A'B'}\epsilon_{C'D'}+\bar{\Psi}_{A'B'C'D'}\epsilon_{AB}\epsilon_{CD} \\
        & \ \ \ +\Phi_{ABC'D'}\epsilon_{A'B'}\epsilon_{CD}+\bar{\Phi}_{A'B'CD}\epsilon_{AB}\epsilon_{C'D'} \\
        & \ \ \ +2\Lambda(\epsilon_{AC}\epsilon_{BD}\epsilon_{A'B'}\epsilon_{C'D'}+\epsilon_{AB}\epsilon_{CD}\epsilon_{A'D'}\epsilon_{B'C'}),
    \end{split}
\end{equation}
where $\Lambda$ and $\Phi_{ABC'D'}=\Phi_{(AB)(C'D')}$ are related to the Ricci scalar and Ricci tensor, respectively, and $\bar{\Phi}_{A'B'CD}$ is the complex conjugate of $\Phi_{ABC'D'}$. We refer to $\Phi_{ABC'D'}$ as  the Ricci spinor. The components of the Ricci spinor can be calculated using the Ricci tensor and the complex null tetrad $\{l,n,m,\bar{m}\}$. There are nine individual components of $\Phi_{ABC'D'}$ \cite{Penrose:1985bww},
\begin{equation}\label{RicciComps}
    \begin{split}
       & \Phi_{000'0'}=-\frac{1}{2}R_{\mu\nu}n^\mu n^\nu, \ \ \ \ \ \Phi_{000'1'}=-\frac{1}{2}R_{\mu\nu}n^\mu m^\nu, \ \ \ \ \ \Phi_{001'1'}=-\frac{1}{2}R_{\mu\nu}m^\mu m^\nu, \\
        & \Phi_{010'0'}=-\frac{1}{2}R_{\mu\nu}n^\mu \bar{m}^\nu, \ \ \ \ \ \Phi_{010'1'}=-\frac{1}{2}R_{\mu\nu}n^\mu l^\nu+3\Pi, \ \ \ \ \ \Phi_{011'1'}=-\frac{1}{2}R_{\mu\nu}m^\mu l^\nu, \\
        &\Phi_{110'0'}=-\frac{1}{2}R_{\mu\nu}\bar{m}^\mu \bar{m}^\nu, \ \ \ \ \ \Phi_{110'1'}=-\frac{1}{2}R_{\mu\nu}\bar{m}^\mu l^\nu, \ \ \ \ \ \Phi_{111'1'}=-\frac{1}{2}R_{\mu\nu}l^\mu l^\nu.\\
    \end{split}
\end{equation}
Here, $\Pi$ is related to the Ricci scalar via $\Pi=\frac{1}{12}R$. The scalar functions $\{\Phi_{000'0'},...,\Phi_{111'1'}\}$ can also be obtained via contractions with $\{o^A,\iota^A\},$ for example (cf. page 22 of \cite{Penrose:1986ca}),
\begin{equation}
    \Phi_{000'0'}=\Phi_{ABC'D'}o^Ao^B\bo^{C'}\bo^{D'}, \ \ \ \ \ \Phi_{110'1'}=\Phi_{ABC'D'}\iota^A\iota^B\bo^{C'}\bi^{D'},
\end{equation}
and so forth. Following the notation of \cite{Penrose:1985bww}, the Ricci scalars are conveniently written as
\begin{equation}
    \begin{split}
        \Phi_{00}\equiv\Phi_{000'0'}, \ \ \ \Phi_{01}&\equiv \Phi_{000'1'}, \ \ \ \Phi_{02}\equiv \Phi_{001'1'}, \\
        \Phi_{10}\equiv\Phi_{010'0'}, \ \ \ \Phi_{11}&\equiv \Phi_{010'1'}, \ \ \ \Phi_{12}\equiv \Phi_{011'1'}, \\
        \Phi_{20}\equiv\Phi_{110'0'}, \ \ \ \Phi_{21}&\equiv \Phi_{110'1'}, \ \ \ \Phi_{22}\equiv \Phi_{111'1'}.
    \end{split}
\end{equation}
Note $\Phi_{ABC'D'}$ is generally \emph{Hermitian}, since $\Phi_{ab}=\bar{\Phi}_{ba}$. 

Using the normalized spinor basis, we can deduce an expansion for the Ricci spinor:
\begin{equation}\label{RicciSpinExpansion}
    \begin{split}
        \Phi_{ABC'D'}&=\Phi_{22}o_Ao_B\bo_{C'}\bo_{D'}-2\Phi_{12}o_{(A}\iota_{B)}\bo_{C'}\bo_{D'}+\Phi_{02}\iota_A\iota_B\bo_{C'}\bo_{D'} \\
        &  -2\Phi_{21}o_Ao_B\bo_{(C'}\bi_{D')}  + 4\Phi_{11}o_{(A}\iota_{B)}\bo_{(C'}\bi_{D')}-2\Phi_{01} \iota_A\iota_B\bo_{(C'}\bi_{D')} \\
        &  +\Phi_{20}o_Ao_B\bi_{C'}\bi_{D'}-2\Phi_{10}\iota_{(A}o_{B)}\bi_{C'}\bi_{D'}+\Phi_{00}\iota_A\iota_B\bi_{C'}\bi_{D'}.
    \end{split}
\end{equation}
As was discussed in the text, the only non-zero Ricci coefficient in Einstein-Maxwell theory is $\Phi_{11},$ thus the Ricci spinor is
\begin{equation}
    \Phi_{ABC'D'}=4\Phi_{11}o_{(A}\iota_{B)}\bar{o}_{(C'}\bar{\iota}_{D')}\, .
\end{equation}

\bibliography{EMwdcrefs}

\end{document}